\documentclass{article}

\usepackage{arxiv}

\usepackage{authblk}        
\usepackage[utf8]{inputenc} 
\usepackage[T1]{fontenc}    
\usepackage{hyperref}       
\usepackage{url}            
\usepackage{booktabs}       
\usepackage{amsfonts}       
\usepackage{nicefrac}       
\usepackage{microtype}      
\usepackage{lipsum}
\usepackage{graphicx}
\graphicspath{ {./images/} }
\usepackage{amsmath}
\usepackage{array}
\usepackage[version=3]{mhchem}
\usepackage[sorting=none, url=false, style=numeric]{biblatex}
\usepackage{multicol}
\usepackage{wrapfig}
\usepackage{float}
\usepackage{lineno}
\usepackage{csquotes}

\usepackage[english]{babel}
\bibliography{references}

\title{Ion correlations explain kinetic selectivity in diffusion-limited solid state synthesis reactions}

\author[1,2]{Vir Karan}
\author[1,2]{Max C. Gallant}
\author[1,2]{Yuxing Fei}
\author[1,2]{Gerbrand Ceder}
\author[1,2]{Kristin A. Persson \thanks{Corresponding Author: kapersson@lbl.gov}}
\affil[1]{Materials Sciences Division, Lawrence Berkeley National Laboratory, Berkeley, CA, 94720}
\affil[2]{Department of Materials Science and Engineering, University of California, Berkeley, CA, 94720}

\date{}

\begin{document}
\maketitle
\begin{abstract}

Establishing viable solid-state synthesis pathways for novel inorganic materials remains a major challenge in materials science. Previous pathway design methods using pair-wise reaction approaches have navigated the thermodynamic landscape with first-principles data but lack kinetic information, limiting their effectiveness. This gap leads to suboptimal precursor selection and predictions, especially for reactions forming competing phases with similar formation energies, where ion diffusion is a critical influence. Here, we demonstrate an inorganic synthesis framework by incorporating machine learning-derived transport properties through "liquid-like" product layers into a thermodynamic cellular reaction model. In the Ba-Ti-O system, known for its competitive polymorphism, we obtain accurate predictions of phase formation with varying \ce{BaO}:\ce{TiO2} ratios as a function of time and temperature. We find that diffusion-thermodynamic interplay governs phase compositions, with cross-ion transport coefficients critical for predicting diffusion-limited selectivity. This work bridges length and time scales by integrating solid-state reaction kinetics with first-principles thermodynamics and spatial reactivity.

\end{abstract}

\newpage

\section{Introduction}\label{sec:intro}
In the past decade, the demand for new inorganic materials to improve energy technologies has led to the emergence of powerful, data-driven materials design. However, realizing in-silico-designed materials through inorganic synthesis remains a challenge, lagging behind organic synthesis due to the absence of a general mechanistic model for solid-state reactions\cite{disalvo_solid-state_1990, schmalzried2008chemical}. Current state-of-the-art atomistic modeling of solid state synthesis describe reaction behavior in terms of bulk thermodynamic properties from high-throughput databases of density functional theory (DFT) calculations such as the Materials Project (MP) \cite{jain_commentary_2013}. Prominent examples include reaction networks~\cite{mcdermott_graph-based_2021} which produce thermodynamically favorable reaction pathways linking products and reactants and active learning algorithms~ \cite{szymanski_autonomous_2023} which propose recipes based on thermodynamics and then updates the recipes according to experimental results.  While thermodynamics defines the possible products, predictions based solely on reaction energetics can be inaccurate — especially for systems with competing phases that have similar formation driving forces\cite{szymanski_quantifying_2024}. In such cases, limited transport of essential constituents may prevent the formation of globally stable products, hindering attainment of thermodynamic equilibrium. Prior attempts to understand such effects in solid-state reactions have led to the use of empirical rate expressions\cite{khawam2006solid, frade1992reexamination}, which fit an effective rate constant from the degree of conversion of the reactants. Although useful, such models can only be fit after-the-fact, and hence cannot be applied to a-priori predict solid-state reaction products.

\begin{figure}
    \centering
    \includegraphics[width=\linewidth]{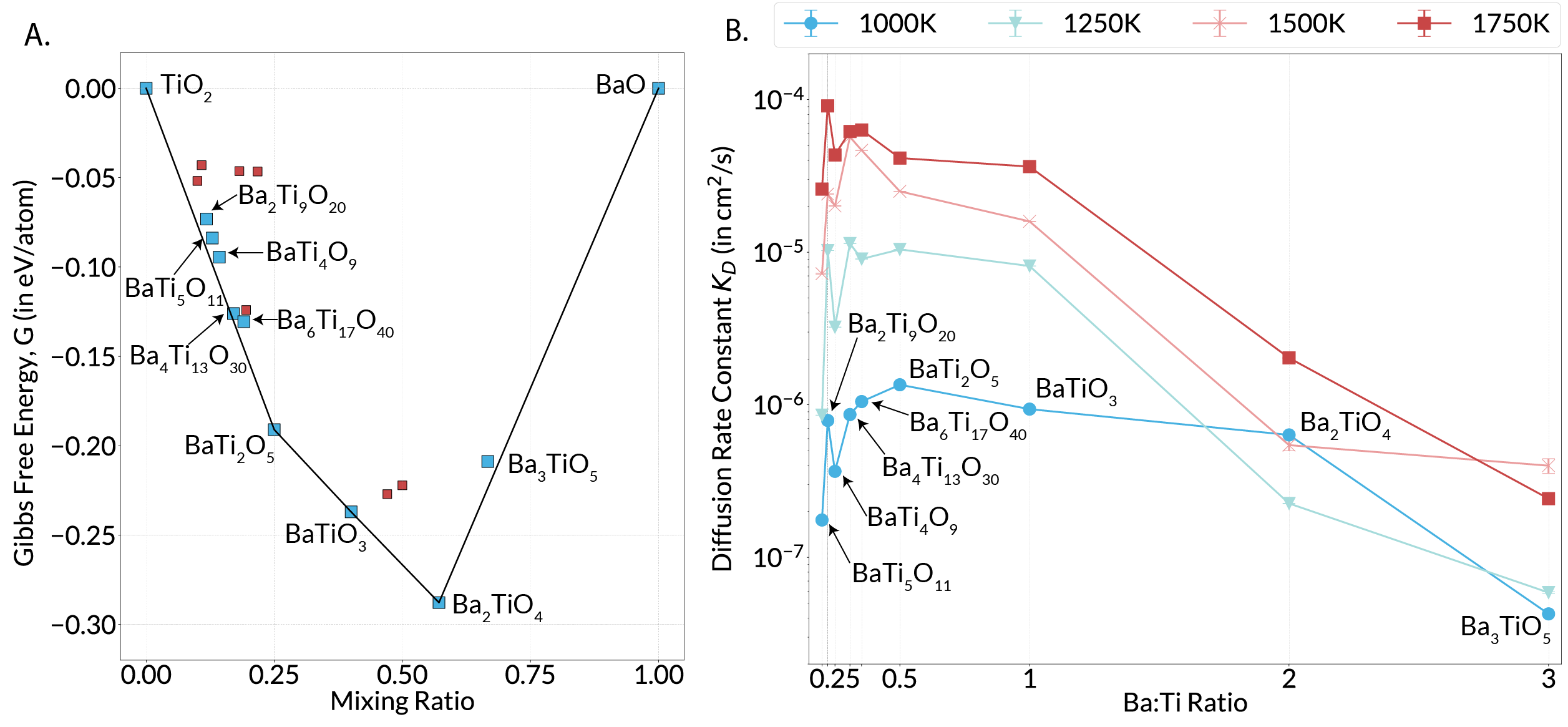}
    \caption{A. The thermodynamic hull of stability for \ce{BaO} and \ce{TiO2} at $600^{\circ}C$, computed using entries from the Materials Project. Finite temperature effects are accounted for using a machine-learned estimator for the vibrational contribution to entropy\cite{bartel_physical_2018}. The blue squares (excepting \ce{Ba3TiO5}) are phases that have been experimentally observed in solid-state reactions, and the red ones are additional phases we consider in this study. The shape of the hull (i.e., the phases and their relative formation energies) remains unchanged for all temperatures we consider in this work; B. The calculated effective diffusion rate constant ($K_D$), which is a measure for the average flux of all ions through the product phase of a reaction, across the \ce{BaO}-\ce{TiO2} reaction interface at a series of temperatures in the range of 1000-1750~K, ordered by increasing Ba:Ti ratio. $K_D$ peaks for phases with intermediate Ba:Ti ratio, and drops significantly in phases with Ba:Ti > 1 at typical synthesis temperatures.}
    \label{fig:Hull+K_D}
\end{figure}

In general, solid-state reaction kinetics can be broken down into nucleation-limited and diffusion-limited regimes \cite{cordova_synthesis_2020}. Although the nucleation-limited regime provides insight into the first phase(s) that forms, which is useful in e.g. thin film synthesis, such information does not conclusively predict the bulk distribution of products in powder reactions, which proceeds by diffusion-controlled transfer of the precursor constituents to the reaction zone. We hypothesize that the synthesis evolution of such systems can be described as an optimization of the $local$ energy under the time-dependent constraint of available ionic fluxes through a defective, liquid-like interphase with the same stoichiometry as candidate, nucleating phases. To showcase our approach and specifically, the effects of diffusive fluxes on product selectivity, we choose the Ba-Ti-O chemical space, which is particularly challenging due to the sheer number of ternary phases which are on or very close to the convex hull of stability (Fig. \ref{fig:Hull+K_D}A). The synthesis of technologically relevant ternaries in this system, e.g. \ce{BaTiO3}, \ce{Ba2Ti9O20}, \ce{BaTi5O11} and \ce{BaTi2O5} have been obtained by reacting primarily \ce{BaCO3} and \ce{TiO2} precursors at differing ratios. In particular, the synthesis of the ferroelectric \ce{BaTiO3} is a well-studied reaction\cite{Beauger1983SynthesisRO, brzozowski2002baco}, with the most common recipe involving the mixing of the binary powders and heating them at temperatures ranging from $1000$ to $1300^{\circ}C$. The reaction generally produces \ce{BaTiO3} as the major product, although notably, \ce{Ba2TiO4} is energetically more favorable at the interface between \ce{TiO2} and \ce{BaO} (Fig. \ref{fig:Hull+K_D}). The target \ce{BaTiO3} is often accompanied by impurities influenced by the synthesis conditions, with \ce{Ba2TiO4} impurities dominating below ~1050 K and higher temperatures ($\approx1200 K$) increasing \ce{BaTiO3} yield but also promoting secondary phases such as \ce{BaTi2O5}.

Previous studies \cite{osti_6480472, brzozowski2002baco}, and in particular recent work by McDermott et al.\cite{mcdermott2023assessing} and Szymanski et al.\cite{szymanski_quantifying_2024} have analyzed selectivity in solid-state synthesis reactions based on first-principles thermodynamic data. Szymanski et al. \cite{szymanski_quantifying_2024} suggested a thermodynamic threshold of 60 meV/atom above which the initial product formation can be reliably predicted. In contrast, for systems with multiple competing phases of comparable driving forces less than the threshold, the initial product was found to be often determined by kinetics. For the Ba-Ti-O system, the formation energy difference between the product with the highest formation driving force (\ce{Ba2TiO4}) and \ce{BaTiO3} and \ce{BaTi2O5} is $\approx $51~meV/atom and $46$ meV/atom, respectively. Hence, we expect synthesis outcomes in this system to be impacted by a temperature-dependent interplay between diffusive fluxes and reaction energies. Indeed, experimentally, the first observed product is generally \ce{Ba2TiO4} \cite{mcdermott2023assessing}, however products with lower driving force are subsequently found, such as \ce{BaTi2O5} and \ce{BaTiO3}.  Using the Ba-Ti-O system as an exacting test case,  we here present a general, predictive framework for solid-state synthesis outcomes that integrates rigorously computed ionic transport from first-principles thermodynamics with a cellular automata, a discrete computational model, where grid cells evolve based on local neighbor interactions. We showcase remarkable agreement between the predicted results, as a function of time and temperature, to four prior, carefully characterized experimental investigations.

\begin{figure}
    \centering
    \includegraphics[width=\linewidth]{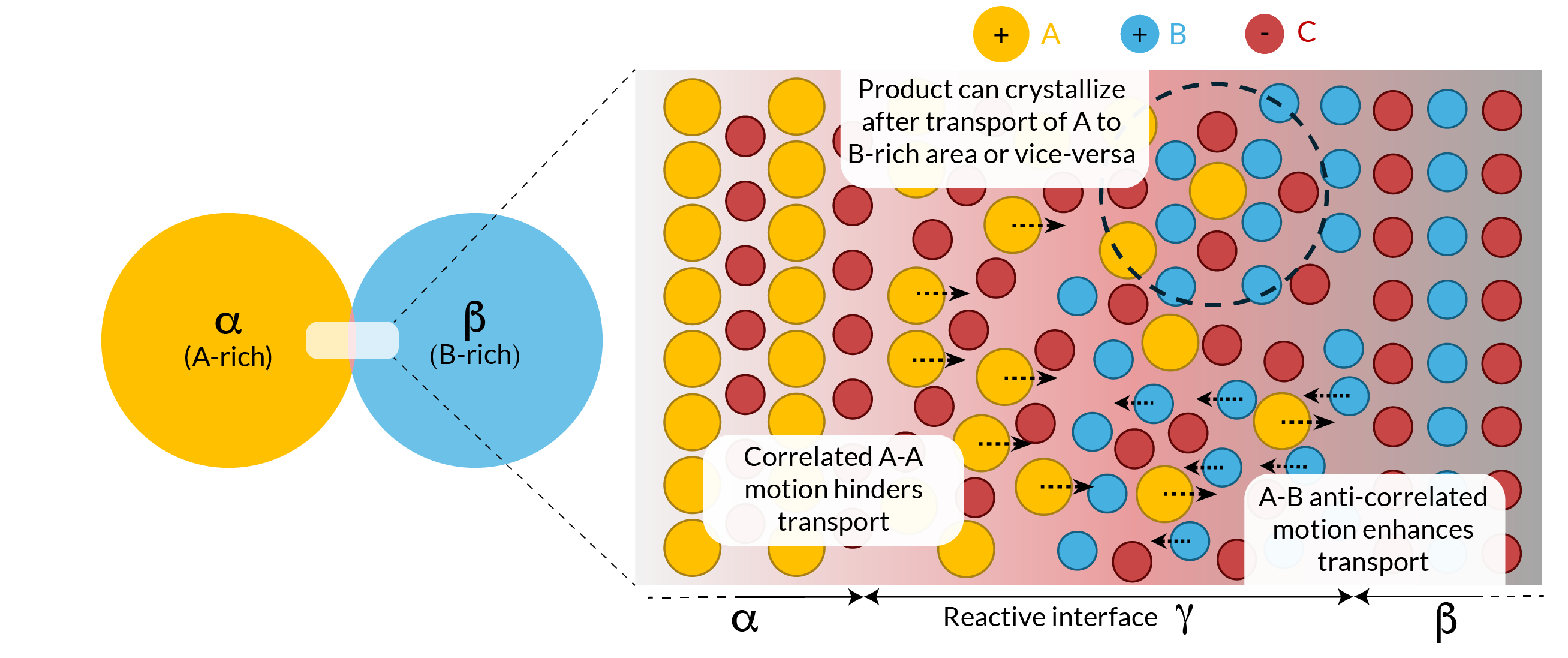}
    \caption{Schematic illustration for a reactive interface between two precursors, one of which is cation A-rich ($\alpha$) and the other is cation B-rich ($\beta$), which react to form an interphase $\gamma$. The diffusion of A towards the B-rich precursor through the disordered/liquid-like region of the interface will determine the local availability of A and B ions, which in turn controls which phases can form and at what rate. }
    \label{fig:toy_model}
\end{figure}




\section{Results}
\label{sec:results}

\subsection{Kinetics in the Ba-Ti-O phase space}
\label{sec:BTO}
We consider two reactants (here \ce{BaO} and \ce{TiO2}) and nine possible solid-state reaction products in the Ba-Ti-O space, marked by blue squares in Fig. \ref{fig:Hull+K_D}. We use \ce{BaO} instead of  \ce{BaCO3}, as it is recognized that the experimental precursor \ce{BaCO3} decomposes to \ce{BaO} at $\approx 1100K$ \cite{buscaglia2008solid}, before any ternary reaction occurs\cite{Beauger1983SynthesisRO}. We calculate the flux of constituent ions from chemical potential difference across the interface (see Fig. \ref{fig:toy_model}) and the transport coefficients of Ba$^{2+}$, Ti$^{4+}$ and O$^{2-}$ using Onsager analyses. These analyses are based on 5 nanosecond MD trajectories generated by machine-learned interatomic potentials (MLIP) trained on AIMD data for each liquid-like, non-crystalline analogue of nine possible products in the Ba-Ti-O system (see Methods~\ref{sec:onsager} for more details). 
  
Using ionic fluxes of both Ba$^{2+}$ and Ti$^{4+}$, effective diffusion rate constants ($K_D$) are estimated for each considered amorphous interphase product (see Methods \ref{sec:ratederivation}). Fig.~\ref{fig:Hull+K_D}B shows $K_D$ for nine possible product compositions across the \ce{BaO} - \ce{TiO2} interface for temperatures 1000-1750~K. Interestingly, above 1000~K, $K_D$ in Ti-rich phases is more than an order of magnitude higher than that in the Ba-rich phases. In the Ti-rich phases, $K_D$ increases by around an order of magnitude with every 250~K rise in temperature, and plateaus to a similar value for most phases at 1750~K. On the other hand, $K_D$ only increases by an order of magnitude in Ba-rich phases (Ba:Ti ratio > 1) on raising temperature by 750K.  

\begin{figure}
    \centering
    \includegraphics[width=\linewidth]{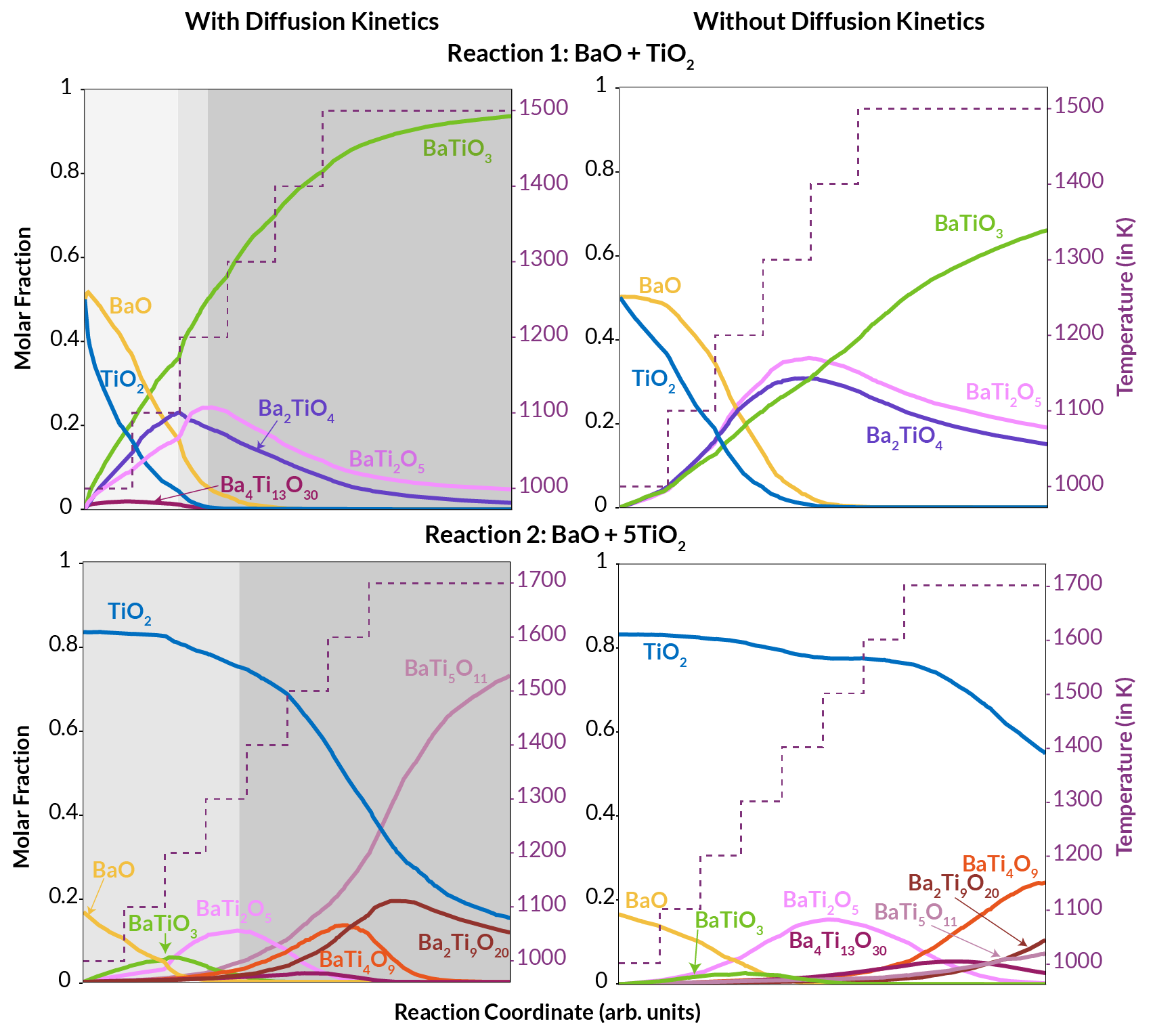}
    \caption{Reaction simulations using \ce{BaO} and \ce{TiO2} precursors with different ratios using ReactCA, informed by both diffusive fluxes and reaction thermodynamics (left) vs only thermodynamics (right) for Reaction 1 (\ce{BaO} + \ce{TiO2}) and Reaction 2 (\ce{BaO} + 5\ce{TiO2}). The shade of gray background distinguishes the different regimes of reaction selectivity: the light gray region signifies the ``Activation-controlled regime'', the darker gray region signifies the ``Kinetics-controlled regime'' and the dark region signifies the ``Thermodynamic-controlled regime''. }
    \label{fig:R_1_2}
\end{figure}

\begin{figure}
    \centering
    \includegraphics[width=\linewidth]{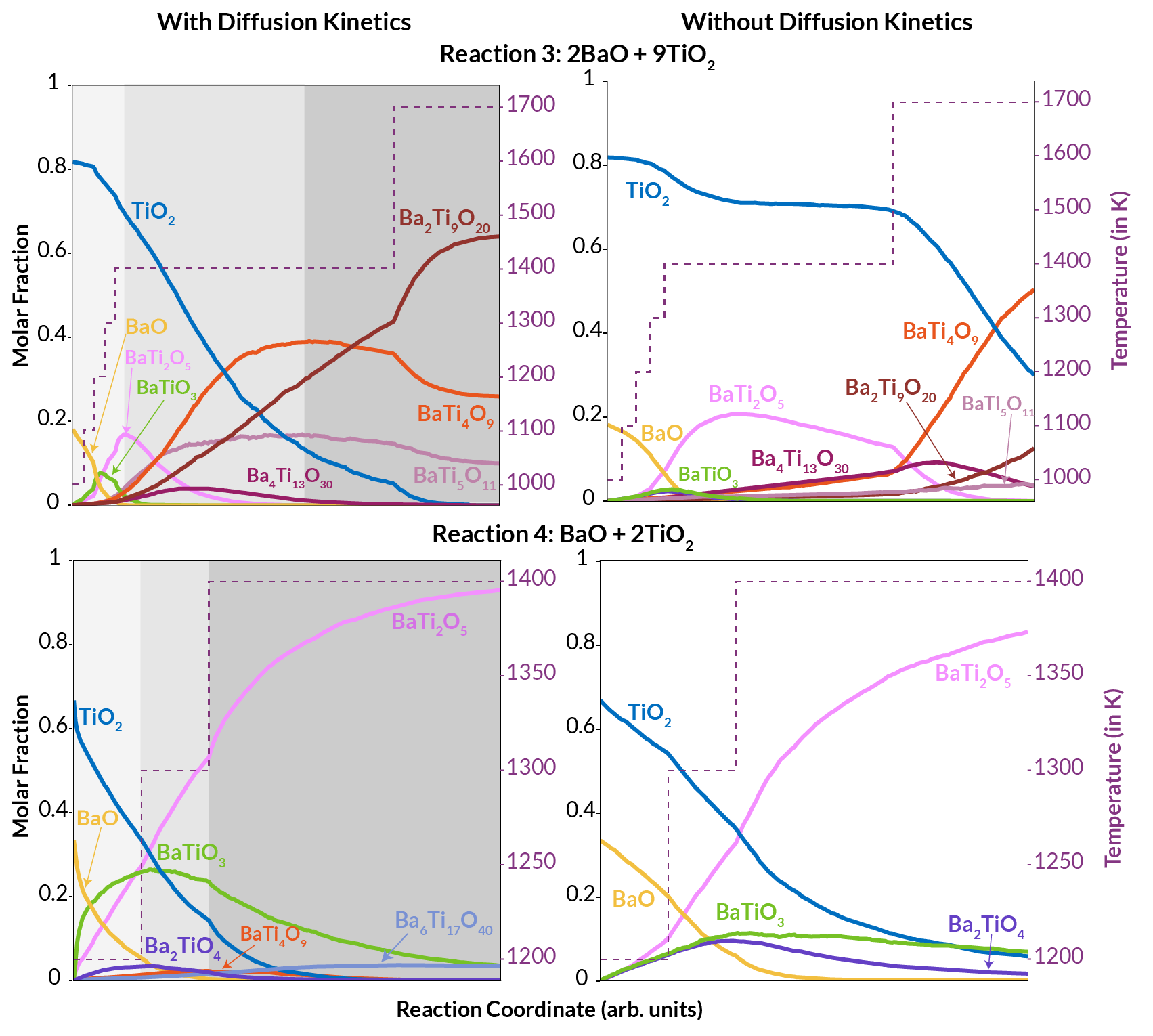}
    \caption{Reaction simulations using \ce{BaO} and \ce{TiO2} precursors with different ratios using ReactCA, informed by both diffusive fluxes and reaction thermodynamics (left) vs only thermodynamics (right) for Reaction 3 (2\ce{BaO} + 9\ce{TiO2}) and Reaction 4 (\ce{BaO} + 2\ce{TiO2}). The shade of gray background distinguishes the different regimes of reaction selectivity: the light gray region signifies the ``Activation-controlled regime'', where kinetics is slow for all compositions, the darker gray region signifies the ``Kinetics-controlled regime'' where kinetics dominates the phase selectivity among thermodynamically competitive phases and the dark region signifies the ``Thermodynamic-controlled regime'', where kinetics is fast for all compositions. }
    \label{fig:R_3_4}
\end{figure}



\subsection{Kinetics-informed Cellular Automaton Simulations of Ba-Ti-O Solid State Synthesis Reactions}

To study the temperature dependence and temporal evolution of the reactions and compare with synthesis experiments reported in McDermott et al.\cite{mcdermott2023assessing}, Wu et al.\cite{wu1988factors}, Beauger et al.\cite{Beauger1983SynthesisRO} and Bierach\cite{osti_6480472}, we use the recently developed cellular automaton simulation framework 'ReactCA'. 
\cite{gallant2024cellular}. ReactCA was designed to simulate solid-state reactions by modeling a 3D grid of cells that evolve based on customizable local rules, incorporating both thermodynamic and kinetic inputs through a proxy for the reaction rate, e.g. a 'scoring' function. Here, we extend the ReactCA framework by allowing the scoring function to depend on the instantaneous growth rate, which is a function of a calculated effective ionic diffusion constant of the amorphous product phase($K_D$) at temperature $T$, a modified thermodynamic driving force $\frac{\Delta G^*}{k_BT}$ and a heuristic for Tammann's rule. \cite{merkle2005tammann} Simplified,  below the Tammann temperature,  reaction rates are low but possible. Above it, rates increase with temperature due to both diffusion and thermodynamic contributions, but at high temperatures, the saturation of diffusion rates shifts the balance in favor of thermodynamically controlled outcomes. We simulate precursor Ba:Ti stoichiometries from 1:5 to 1:1 using the same heating profiles as the experiments (Fig. \ref{fig:R_1_2}-\ref{fig:R_3_4})~\cite{bondioli_kinetic_1998, mcdermott2023assessing, osti_6480472, wu1988factors}. Simulations using a scoring function which excludes the effective diffusion rates, i.e., reaction rates based only on thermodynamics and Tamman's rule, are shown in the same figure for comparison. 

Reaction 1, shown in the top panels of Fig. \ref{fig:R_1_2} represents the prototypical 1:1 \ce{BaO} to \ce{TiO2} synthesis process, involving heating to $1500K$ followed by a brief annealing period \cite{bondioli_kinetic_1998, mcdermott2023assessing}, resulting in \ce{BaTiO3} as the predominant product. In our simulation, the first products to form at low temperatures are \ce{Ba2TiO4} and \ce{BaTiO3}. This agrees excellently with experimental results showing \ce{Ba2TiO4} as the primary impurity (up to 27 mol\%) \cite{mcdermott2023assessing, Beauger1983SynthesisRO}  when the reaction was performed between 1000-1050K. At higher temperatures, we find that \ce{Ba2TiO4} is consumed in favor of \ce{BaTi2O5}, which emerges as the major impurity past 1200~K, in agreement with McDermott et al. \cite{mcdermott2023assessing} and Bierach \cite{osti_6480472}. With continued heating, \ce{BaTi2O5} is mostly consumed but persists as the primary  impurity ($<5 mol\%$) at the conclusion of the reaction. Indeed, Bondioli et al. \cite{bondioli_kinetic_1998} found the reaction is driven to completion when performed for >100 minutes at 1300K, giving almost phase-pure \ce{BaTiO3}. Performing the same simulation without incorporating diffusion rates yields a qualitatively different outcome. Results from simulations that incorporate only the thermodynamics-based data as well as Tamman's rule show a mixture of \ce{BaTiO3} ($\approx 70 mol\%$), \ce{Ba2TiO4}($\approx 15 mol\%$), and \ce{BaTi2O5} ($\approx 20 mol\%$) after annealing at 1500K, with \ce{Ba2TiO4} and \ce{BaTi2O5} amounts peaking at similar temperatures. Notably, \ce{BaTiO3} accounts for only 70\% of the product, in stark contrast to the nearly phase-pure \ce{BaTiO3} observed experimentally.

Reaction 2 (bottom panels of Fig. \ref{fig:R_1_2}) follows the experiments conducted by O'Bryan and Thomson \cite{o1975preparation} where a 1:5 ratio of \ce{BaO} and \ce{TiO2} was reacted at a series of temperatures ($1250-1500K$). At low temperatures ($< 1300K$) our simulation suggests the formation of small amounts of \ce{BaTiO3} ($< 10 mol\%$) and \ce{BaTi2O5} ($\approx10 mol\%$) which are rapidly consumed in favor of \ce{BaTi5O11}, \ce{Ba2Ti9O20} and an intermediate: \ce{BaTi4O9}. \ce{BaTi4O9} reaches a maximum of $\approx 15 mol\%$ at 1500K before its consumed in favor of the high temperature products \ce{BaTi5O11} and \ce{Ba2Ti9O20}, together with remaining precursor \ce{TiO2}. Interestingly, only \ce{BaTi5O11}, \ce{BaTi4O9} and unreacted rutile (\ce{TiO2}) were present in the final product distributions of O'Bryan and Thomson. Since they do not specify the annealing or reaction times, we suggest that their experiment maps onto the early stages of our simulation, before a considerable amount of \ce{Ba2Ti9O20} forms. Continuing the reaction longer at a higher temperature would have promoted the formation of \ce{Ba2Ti9O20}, which also qualitatively agrees with the experiments conducted in Ref~\cite{obryan1983ba2ti9o20} where \ce{Ba2Ti9O20} forms at the expense of \ce{BaTi4O9} and \ce{TiO2} above $1400K$. 

Wu et al.\cite{wu1988factors} reacted a 2:9 ratio of \ce{BaO} and \ce{TiO2} (Reaction 3), by rapidly increasing the temperature to $\approx$1400K, followed by annealing for 3 hours.  The sample was then sintered at approximately 1700K for 6 hours, yielding \ce{Ba2Ti9O20} as the primary product along with impurities of \ce{BaTi5O11} and \ce{BaTi4O9}. Notably, \ce{Ba2Ti9O20} was predominantly formed at the expense of \ce{BaTi5O11} and \ce{BaTi4O9}. Wu et al. also observe the relative rates of formation of the three phases as follows: \ce{BaTi5O11} > \ce{BaTi4O9} > \ce{Ba2Ti9O20}. In our simulation (see Fig. \ref{fig:R_3_4}), small amounts of \ce{BaTiO3} and \ce{BaTi2O5} are initially formed, but quickly consumed in favor of \ce{Ba2Ti9O20}, \ce{BaTi5O11} and \ce{BaTi4O9}. In particular, our results successfully reproduce the experimentally observed trend in initial formation rates, following the sequence: \ce{BaTi5O11} > \ce{BaTi4O9} > \ce{Ba2Ti9O20}. Furthermore, and in agreement with experiments, the growth of \ce{BaTi5O11} and \ce{BaTi4O9} stagnates during the 1400K anneal and finally declines above 1400K, whereas  \ce{Ba2Ti9O20} is continually produced at a positive rate. Wu et al. also detected trace quantities of \ce{Ba4Ti13O30} and some unreacted rutile (\ce{TiO2}) when the reaction was performed below 1400K, which are also observed in our simulations. 

Zhu et al.\cite{zhu2010formation} provided a recipe to synthesize phase pure \ce{BaTi2O5} by annealing a 1:2 \ce{BaO}:\ce{TiO2} precursor ratio at three successive temperature steps: $\approx 1200K$, $1300K$ and $1400K$, which we simulate through Reaction 4 (Fig. \ref{fig:R_3_4}). After the first step, their sample primarily contained \ce{BaTiO3} and some amount of \ce{BaTi2O5}. After the second annealing step, the \ce{BaTiO3} content was reduced and \ce{BaTi2O5} increased proportionately. After the third annealing step, close to phase pure \ce{BaTi2O5} was observed with \ce{BaTiO3} and \ce{Ba6Ti17O40} appearing as impurity phases. Upon longer heating time above $1400K$, \ce{BaTiO3} and \ce{Ba6Ti17O40} (the equilibrium phases above $1473K$) increase in phase fraction \cite{zhu2010formation}. Our simulations show excellent agreement; we show a rapid formation of \ce{BaTiO3} as the majority phase in the first annealing step, which is then overtaken by \ce{BaTi2O5} in the second annealing step and we recover phase-pure \ce{BaTi2O5}, with trace (<5\%) impurities of \ce{Ba6Ti17O40} and \ce{BaTiO3}, as the final products. 

Finally, in their study of the formation of \ce{BaTiO3} by the conventional solid-state reaction in air, Bierach \cite{osti_6480472} mentions the formation of \ce{BaTi4O9} and \ce{Ba4Ti13O30} phases concurrently with \ce{BaTiO3} in the early stages of the reaction. They attribute the presence of these impurity phases to kinetic effects, i.e., phases that, despite having only modest thermodynamic driving forces for formation, can nucleate and grow due to their rapid diffusion kinetics. Our simulations align well with experimental observations, showing the formation of these phases alongside \ce{BaTiO3}, with their occurrence depending on the overall system composition. Ti-rich systems predominantly produce trace amounts of \ce{BaTi4O9} (Reaction 4), while more compositionally balanced systems yield trace amounts of \ce{Ba4Ti13O30} (Reaction 1).


\section{Discussion}
\label{sec:discussion}




Our approach, which incorporates rigorously calculated ionic diffusion coefficients for a liquid-like reactive interphase and integrates them through an effective diffusion rate constant ($K_D$), significantly improves the capability to a-priori predict solid-state synthesis pathways and outcomes as a function of temperature and time. Fig. \ref{fig:R_1_2}~-~\ref{fig:R_3_4} match well with available carefully characterized experiments, in terms of reproducing the right products and intermediates as a function of temperature profile and time. Notably, when performing the same simulations using only thermodynamic reactivity data and Tamman's rule, the same framework often predicts the correct major product, while failing to accurately capture the intermediate products and major impurities, which are either absent or incorrectly identified.

\begin{figure*}
    \centering
    \includegraphics[width=\linewidth]{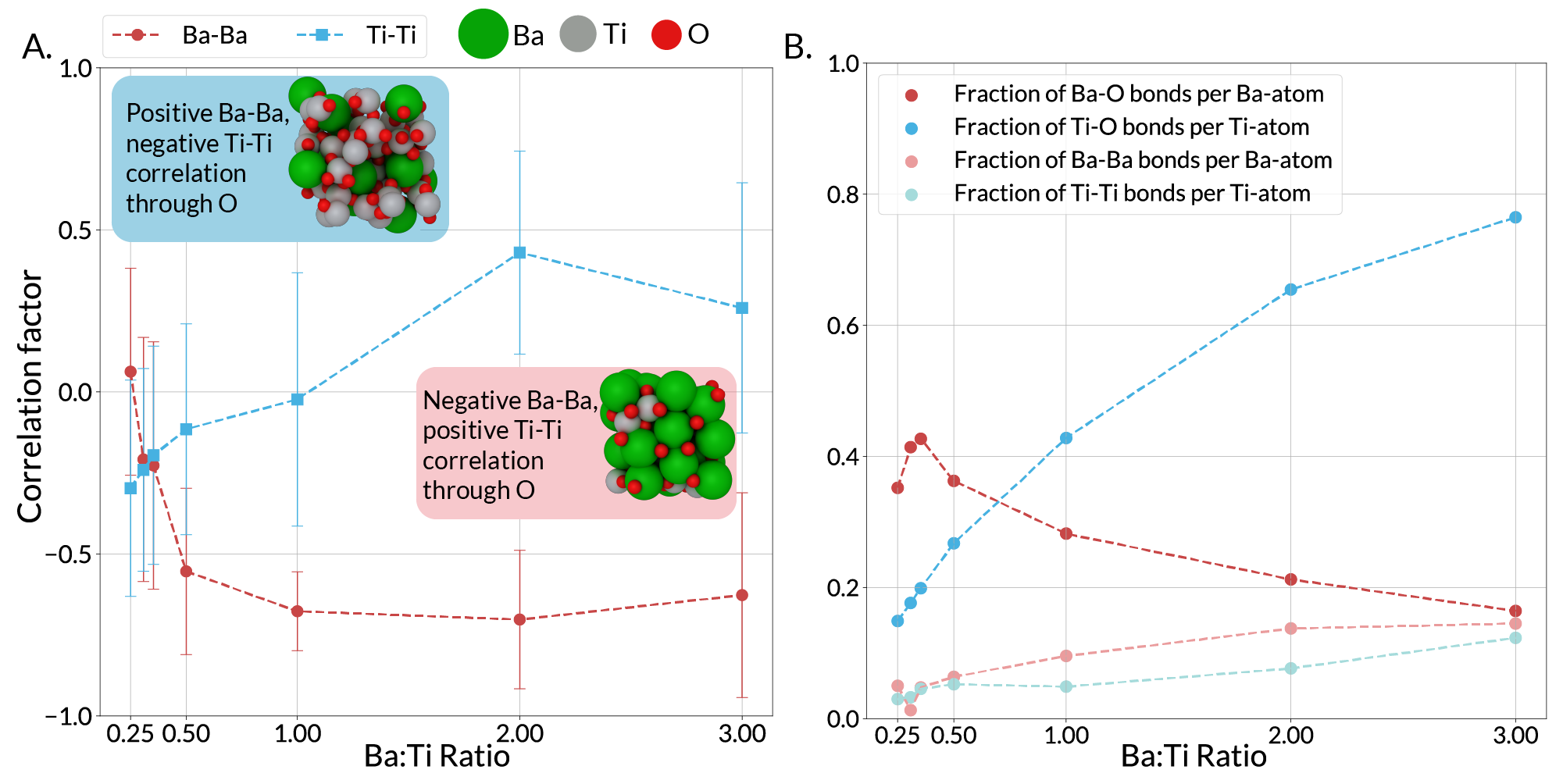}
    \caption{A: Correlation factors (ratio of distinct to self transport coefficient) for Ba (in red) and Ti (in blue), B: Local coordination environment analysis of liquid-like interphases, as a function of Ba:Ti ratio in the phase. On increasing the Ba-content in the phase, a cross-over point exists after which motifs of \ce{TiO_x} clusters dominate with an increased Ba-Ba interaction through a bridging O (evidenced by the higher correlation factor and number of Ba-Ba bonds). Correspondingly, the distinct ion correlation of Ba-Ba becomes negative, implying repulsive behavior between neighboring $Ba^{2+}$ ions leading to a reduction in mobility of $Ba^{2+}$ ions.}
    \label{fig:clusters}
\end{figure*}

The strong agreement between experiments and ab-initio predictions supports the hypothesis that transport through an amorphous interphase matching the target phase approximates experimental conditions well. Using this insight, we analyze the impact of different amorphous compositions on diffusion rates, which is expected to exhibit strong ion correlations.\cite{cheng2022materials} To showcase the impact of correlated ionic motion, i.e., how the movement of one ion in the solid is influenced by the position and motion of nearby ions, we plot the calculated distinct ion Onsager transport coefficients, normalized by the self-diffusion part of the correlation function for the cations ($Ba^{2+}$ and $Ti^{4+}$). Fig. \ref{fig:clusters} depicts the normalized distinct Ba-ion and Ti-ion transport coefficients, in comparison to the number of Ba-O, Ti-O, Ba-Ba and Ti-Ti bonds present~\cite{zimmermann2020local, ong2013python} in the liquid-like interphasial structures for each composition. The data reveals a strong trend between the degree of correlated motion between the $Ba^{2+}$ and $Ti^{4+}$ and the local coordination environment. On increasing the Ba:Ti ratio in the system, the $Ti^{4+}$  environment becomes dominated by O$^{2-}$ which bridge other $Ti^{4+}$, leading to higher fraction of \ce{TiO_x} clusters. This in turn leads to an increase in the number of Ba-Ba bonds per Ba-atom (see Fig. \ref{fig:clusters}) and a higher degree of Ba-Ba correlated motion. The observation aligns with the computed distinct ion correlations: in Ba-rich phases, Ba$^{2+}$ ions show negative correlations, indicating mutual repulsion under a chemical potential gradient, while $Ti^{4+}$ ions exhibit smaller positive correlations, suggesting a weak tendency to cluster. Thus, the net  Ba$^{2+}$ movement slows compared to self-diffusion, reducing its mobility due to cross-correlated motion in Ba-rich phases. A similar, but weaker, effect occurs in Ti-rich phases, causing the diffusion rate constant to peak at an intermediate Ba:Ti ratio (Fig. \ref{fig:Hull+K_D}B). To demonstrate how the correlated motion between the Ba-ions and Ti-ions affects selective phase formation during synthesis, we compare the simulations with and without the cross-ion effects i.e., using only the self-diffusion estimates for ionic  mobility (S.I Fig. S10). While the results are qualitatively similar to Fig. \ref{fig:R_1_2}~-~\ref{fig:R_3_4}, there are key differences, primarily at the initial stage when reactions involve more \ce{Ba}-rich phases. For example,  Reaction 1 does not form the experimentally verified \ce{Ba4Ti13O30} intermediate~\cite{osti_6480472}. In Reaction 3, adding cross-ion fluxes reduces the amount of initial Ba-rich intermediates which agrees with the observations of Wu et. al.\cite{wu1988factors}. Furthermore, in Reaction 3, only \ce{Ba2Ti9O20} and \ce{BaTi4O9} are experimentally observed in the final product distribution~\cite{wu1988factors}, a trend that is more accurately captured in the simulation when cross-transport terms are included. When only the self-transport terms are considered, the simulation overestimates the amount of \ce{Ba2Ti9O20} and underestimates the \ce{BaTi5O11} and vice-versa in Reactions 2 and 3 respectively. We conclude that while self-diffusion terms capture most kinetic effects, incorporating cross-ion fluxes is crucial to accurately capture relative phase formation. These cross-effects are expected to be even more significant in reactions like ion-exchange or metathesis, involving multiple anion groups.~\cite{neilson_modernist_2023}.

Furthermore, by analyzing the difference between the simulations based on thermodynamic data only and thermodynamics combined with kinetic information, we can identify three different temperature regimes for reaction selectivity in the Ba-Ti-O system: the Activation-controlled, the Kinetics-controlled and the Thermodynamic-controlled regimes. At low temperatures (below $\approx1100K$ for the Ba-Ti-O system), where diffusion kinetics is slow across all products, phases that align reasonably with the precursor ratio and which exhibit substantial reaction energies $\Delta G_{rxn}$ are favored (Activation-controlled regime). At intermediate temperatures ($1100K$<T<$1700K$),  diffusion kinetics favor the formation of certain compositions among the possible phases (Kinetics-controlled regime). At high temperatures (T>$1700K$), close to the melting point of the system, kinetics is fast for all ionic species, and the mixture of phases closest to the globally set composition on the thermodynamic hull form to establish thermodynamic equilibrium. For example, \ce{BaTiO3} and \ce{BaTi2O5}, which exhibit large negative formation energies on the Ti-rich side of the hull and relatively faster kinetics (see Fig. \ref{fig:Hull+K_D}), tend to form at lower temperatures (Reaction 2 and 3). Raising the temperature above 1700K, corresponding to the thermodynamic regime, promotes the formation of \ce{BaTi5O11} and \ce{Ba2Ti9O20} which is the thermodynamically most favorable product mixture for the given precursor ratio. 


Finally, we comment on the impact of structural similarity in solid state synthesis. Several prior works \cite{aykol_rational_2021, wu1988factors, szymanski_quantifying_2024} argue that structural similarity or templating between the precursors, intermediates and products promote the formation of specific phases. Indeed, Wu et al. \cite{wu1988factors} hypothesized that \ce{Ba2Ti9O20} forms above $1373K$ only when the intermediate \ce{BaTi5O11} provides a lowering of interfacial and strain energies by templating of the \ce{Ba2Ti9O20} phase onto \ce{BaTi5O11}. Here we reproduce the product distribution of  Wu et al. without incorporating any structural similarity effects. Specifically, we find similar relative rates of formation for the three ternary phases in the final product distribution (\ce{BaTi5O11} > \ce{BaTi4O9} > \ce{Ba2Ti9O20}) during the early stages of Reactions 2 and 3 (Fig. \ref{fig:R_1_2}~-~\ref{fig:R_3_4}) and that, importantly, the \ce{Ba2Ti9O20} phase is only thermodynamically stable above $1373K$, as indicated by the \ce{BaO}-\ce{TiO2} phase diagram \cite{lee_2007_phasediagram}. The modest driving force for its formation (Fig. \ref{fig:Hull+K_D}) accounts for its appearance only at temperatures above $1373K$ in both experiments by Wu et al. \cite{wu1988factors} and our simulations, as well as its absence in the shorter reaction times employed by others \cite{mcdermott2023assessing}. To generally analyze structural effects in the Ba-Ti-O system, we included the nucleation rate computed using the approach by Aykol et al.~\cite{aykol_rational_2021} Their metric shows that \ce{BaTiO3} and \ce{BaTi2O5} exhibit the lowest nucleation barriers and should dominate at low to moderate temperatures (see S.I.). However, \ce{Ba2TiO4} often forms as a low-temperature intermediate in both experiments and simulations, which this approach cannot explain.

In conclusion, using the Ba-Ti-O system as a case study, we present a general simulation framework for solid state synthesis combining rigorously computed transport coefficients with first-principles thermodynamics. We show how ionic transport through liquid-like product layers governs selectivity in diffusion-limited reactions, with simulations of four independent reactions matching experimental results. While self-diffusion terms dominate kinetic effects, cross-ion correlations are essential for explaining intermediate phases and product fractions as a function of time and temperature. We identify three regimes of reaction selectivity: at low temperatures, products with the largest thermodynamic driving force; at intermediate temperatures, interplay between kinetics and thermodynamics; and at high temperatures, where kinetics are similar for all ions, products align with the global thermodynamic composition. This work establishes a foundation for predictive, mechanistic models and digital twins for solid-state synthesis.

\newpage

\section{Methods}
\label{sec:methods}
To capture ion dynamics effects in dense liquid-like or non-crystalline phases, we adapt the framework by Fong et al.\cite{fong2020onsager, fong_transport_2020} to obtain both self as well as correlated diffusive transport coefficients in the form of the Onsager transport matrix. Robust estimates of transport coefficients are obtained from nanosecond long molecular dynamics trajectories, using an atomic cluster expansion (ACE) based machine learning interatomic potential (MLIP) \cite{drautz2019atomic}, trained on 75 ab initio MD trajectories from non-crystalline systems. Resulting kinetic data, as well as thermodynamic data from the Materials Project are implemented in a recently developed cellular automaton (ReactCA) simulation framework \cite{gallant2024cellular} which allows for describing both the spatial and temporal phase evolution over the course of a prescribed solid state reaction.  Below, we describe the components of the simulation framework in more detail. 

\subsection{Accelerating ab-initio molecular dynamics using the Atomic Cluster Expansion (ACE)}
\label{sec:acetraining}

We use the MPMorph \cite{aykol_thermodynamic_2018, zheng2024ab} workflow, as implemented in \texttt{atomate2} \cite{ganose_atomate2_2024} to generate amorphous configurations for training ACE. Sample random amorphous structures of the desired composition are first generated using \texttt{packmol}\cite{packmol}. The MPMorph workflow begins by adjusting the volume of this structure to 0.8 and 1.2 times the initial volume, followed by a 4 picosecond NVT AIMD run to determine the equation of state at the given temperature. A trial volume is then selected based on the equation of state for another 4 picosecond NVT AIMD run to check for energy convergence. If the energy converges, a 20 picosecond AIMD "production" run is performed to equilibrate the structure at this volume. If not, the workflow continues to iteratively rescale the volume until energy convergence is achieved. Once a converged volume is found, it is used for the 20 picosecond production run. The workflow is run on high-performance computing resources using Jobflow \cite{rosen2024jobflow} and FireWorks \cite{jain2015fireworks}. 

Compositions according to \ce{BaTiO3}, \ce{Ba2TiO4}, \ce{BaTi2O5}, \ce{BaTi5O11} were run at 1000, 1250 and 1500K each, and frames were sampled every 100 femtosecond to ensure sufficient variety in the sampled configurations. The sampled frames were used in addition to Materials Project data in the Ba-Ti-O chemical space to train ACE potentials using the \texttt{pacemaker} \cite{lysogorskiy2021performant, bochkarev2022parameterization} python package. The ACE potential is parameterized with 500 basis functions per element, considering neighborhood lists within 5$\mathring{A}$. Following Ref ~\cite{erhard2024modelling}, a higher order Finnis-Sinclair embedding function is used. Training is performed by hierarchically growing the potential using a power-order based ``ladder-fitting'' approach \cite{bochkarev2022parameterization} with a higher weight on the loss contribution from forces (99\%) for 2000 iterations. Once convergence of the training was obtained, the potential was fine-tuned by reducing the force loss weight to 90\% for another 2000 iterations while keeping the shape and complexity of the potential fixed. 

Active learning was performed using the extrapolation grade and the high-temperature MD approach described in Ref ~\cite{lysogorskiy2023active}. We generate randomly packed structures using \texttt{packmol} for compositions corresponding to \ce{BaTi5O11}, \ce{Ba2Ti9O20}, \ce{BaTi4O9}, \ce{Ba4Ti13O30},\ce{Ba6Ti27O40}, \ce{BaTi2O2}, \ce{BaTiO3}, \ce{Ba2TiO4} and \ce{Ba3TiO5} as input for ACE-based NVT MD simulations for 100 picoseconds (100,000 steps) at 2000~K. We employ the extrapolation grade, which quantifies the deviation of the test configurations sampled through ACE-MD from those encountered in the training data to filter out structures that are sufficiently different from the training data. During the high-temperature MD simulations, all configurations encountered by the model with an extrapolation grade > 5 are considered for active learning, and the MD simulation is terminated if the model encounters a configuration with extrapolation grade exceeding 100. A subset of the collected structures is determined using the D-optimality criterion \cite{lysogorskiy2023active} to ensure diversity in the active learning data, and DFT single-point calculations are performed to compute the DFT energy and forces. The trained potential is then fine-tuned on the active learning data for an additional 1000 iterations, and this active learning loop is repeated 4 times. In total, including the data acquired during active learning, the training dataset consisted of 4,628 structures containing 492,216 atoms.

It is important to highlight that the objective here is not to develop a potential capable of addressing a broad variety of tasks (such as crystal relaxation, property calculations etc) within the Ba-Ti-O chemical space, as has been successfully achieved with simpler systems. Specifically, the goal of the trained potential is to obtain longer-time NVT MD trajectories with similar to DFT accuracy on the amorphous ``liquid-like'' configurations. With this purpose in mind, benchmarks for the quality of the trained model are discussed in the S.I. 

\subsection{Onsager transport framework for a solid-state reaction interface}
\label{sec:onsager}

In the framework of linear irreversible thermodynamics, the diffusive flux induced in an ion $i$ due to a driving force can be written as \cite{fong2020onsager, fong_transport_2020}:

\begin{equation}
\label{eqn: fluxeqn}
    J_i = - \Sigma_j L_{ij}\nabla \tilde{\mu}_j
\end{equation}

This expression considers the effect of driving forces ($\{\nabla \tilde{\mu}_j\}$) on all ions present in the system to the flux experienced by ion $i$ through the Onsager transport matrix ($\mathbf{L} = [L_{ij}]$). The transport coefficient, $L_{ij}$ is then computed from an MD trajectory using the differential form of the Green-Kubo relations \cite{fong_transport_2020}:
\begin{equation}
    L_{ij} = \frac{1}{6k_BTV} lim_{t \rightarrow \infty} \frac{d}{dt}<\Sigma_\alpha[\mathbf{r}^i_\alpha(t)- \mathbf{r}^i_\alpha(0)] \cdot \Sigma_\beta[\mathbf{r}^j_\beta(t)- \mathbf{r}^j_\beta(0)]>
\end{equation}

Here V and T are the volume and temperature of the system, $\mathbf{r}^i_\alpha(t) - \mathbf{r}^i_\alpha(0)$ is the displacement of the $\alpha^{th}$ particle of specie $i$ at time $t$. The self transport coefficient, $L_{ii}^{self}$, can be computed using a similar relation:
\begin{equation}
    L_{ii}^{\text{self}} = \frac{1}{6k_BTV} lim_{t \rightarrow \infty} \frac{d}{dt}<\Sigma_\alpha[\mathbf{r}^i_\alpha(t) - \mathbf{r}^i_\alpha(0)]^2>
\end{equation}

The self-transport coefficient of an ion $i$ is related to the Nernst-Einstein estimate of diffusivity, also called the self diffusion coefficient ($D_i$) through the following relation:

\begin{equation}
    L_{ii}^{\text{self}} = \frac{D_{i}c_{i}}{k_BT}
\end{equation}
where $c_i$ is the concentration of specie $i$. This expression is exact for dilute systems, wherein diffusion is ideal. 
Practically, $L_{ij}$ is the slope of a linear fit (the diffusive regime) of the time correlation function, $<\Sigma_\alpha[\mathbf{r}^i_\alpha(t) - \mathbf{r}^i_\alpha(0)] \cdot \Sigma_\beta[\mathbf{r}^j_\beta(t)- \mathbf{r}^j_\beta(0)]>$, with time $t$. Hence, a linear regime of atleast 200 picoseconds (200,000 steps) is used in the time correlation function vs time plot to compute all the transport coefficients. More details on how the transport coefficients are fit, as well as other caveats when using this theory on solid-state reactions is given in the S.I.

\subsection{Deviations in diffusion due to correlated ion movement}

The deviation in effective diffusion coefficients can be studied through the cross-ion correlations, which are quantified by the off-diagonal terms of $\mathbf{L}$, and the distinct ion correlations, which are computed as:
\begin{equation}
    L_{ii}^{\text{distinct}} = L_{ii}^{\text{self}} - L_{ii}
\end{equation}
In this work, we focused on the effect of correlations amongst the cation pairs: Ba-Ba and Ti-Ti, through the corresponding distinct ion correlation terms. To compare the correlation terms between phases, we calculate the distinct correlation normalized by the self part of the correlation function:
\begin{equation*}
    f_i = \frac{L_{ii}^{\text{distinct}}}{L_{ii}^{\text{self}}} = \frac{L_{ii}}{L_{ii}^{\text{self}}} - 1
\end{equation*}
The distinct correlation typically exhibits values between -1 to +1, with values close to 0 implying uncorrelated motion between ions of type $i$. Additionally, we analyze the local coordination environment within the amorphous phases by counting all geometrically feasible bonds for each ion within a cutoff radius of $5\mathring{A}$. This approach captures spatial correlations or bonding interactions within both the first and second coordination shells. Both shells are included because cations typically interact through bridging O-ions. This enumeration is performed using the CrystalNN algorithm implemented in pymatgen \cite{zimmermann2020local, ong2013python}.

\subsection{Estimating reaction kinetics from atomistic transport}
\label{sec:ratederivation}
The amorphous, liquid-like structures generated as outlined in Sec. \ref{sec:acetraining} serve as inputs for NVT molecular dynamics simulations using the ACE potential. These simulations are conducted for 5 nanoseconds with one femtosecond time step at temperatures of 750 K, 1000 K, 1250 K, 1500 K, and 1750 K, employing the Langevin thermostat with a friction factor of 0.01 $\text{femtosecond}^{-1}$. To obtain better statistics on the computed transport coefficients, five frames from the production run of the ACE-based MPMorph flow are used as starting configurations for the NVT MD runs, leading to five replicate MD runs for each composition. 

Following Schmarlzreid et al. \cite{schmalzried_build-up_1986}, as a first order, we approximate the driving forces $\nabla \tilde{\mu}_j$ in equation \ref{eqn: fluxeqn}, which are electro-chemical potential gradients, by the bulk chemical potential gradient for specie $j$, $\nabla \mu_j$. These gradients are obtained through the relevant chemical potential diagram for the chemical system, which is built using pymatgen \cite{ong2013python} with data for all crystalline phases from the Materials Project. We approximate the gradient in chemical potential for specie $j$ between the precursors (or in general the reactants) of a solid-state reaction by the shortest distance on chemical potential diagram between domains corresponding to the precursors (reactants), say $\alpha$ and $\beta$, divided by the thickness of the product layer ($h$ in Fig. S7 in the S.I):
\begin{equation}
\label{eqn:mu_approx}
    \nabla \tilde{\mu}_j \approx \nabla \mu_j = \frac{\min(\mu_j^\alpha - \mu_j^\beta)}{h}
\end{equation}
This approach ensures that when the phases are in thermodynamic equilibrium, the reaction driving force and the resulting flux are zero\cite{neilson_modernist_2023}. The effective ``diffusion rate constant'' ($K_D$) is defined as the following:
\begin{equation}
\label{eq:k_d}
    K_D = \Sigma_i \Sigma_j \frac{|L^{\gamma}_{ij}V^{\gamma} \cdot \text{min}(\mu_i - \mu_j)|}{n^{\gamma}_i}
\end{equation}
where $V^{\gamma}$ is the molar volume and $n^{\gamma}_i$ is the number of atoms of specie $i$ in product phase $\gamma$. We assume a core shell model for the formation of the product (derivation shown in the S.I.), with growth of $\gamma$ supported by transport of Ba$^{2+}$ and Ti$^{4+}$ across the interface. Notably, the growth rate varies with the thickness of the product phase, and decreases as the reaction proceeds. For a purely diffusion-limited case for the geometry shown in S.I Fig. S7, the rate equation can be solved to give \cite{lu1998kinetic, dheurle_theoretical_1995}:
\begin{equation}
    1 - (1 - y)^{1/3} = \frac{\sqrt{2K_Dt}}{r_0}
\end{equation}
where y ($1\geq y\geq0$) is the degree of completion of the reaction and $r_0$ is the radius of the precursor powder particle \cite{lu1998kinetic}, reminiscent of the Jander equation which is commonly used to fit kinetic models to solid-state reaction data \cite{khawam2006solid,frade1992reexamination}. In this study, we assume that powder particles are well-mixed and uniform in size, allowing us to analyze the kinetic feasibility of a solid-state reaction using the term $\frac{\sqrt{K_D}}{r_0}$, where $r_{BaO} = r_{TiO_2} = r_0$. The parameter $r_0$ is set uniformly for all reactions in our simulations. Specifically, we assign $r_0$ a value of $10^{-1}$ microns, which is representative of particle sizes commonly used in solid-state synthesis.

We emphasize that this formulation is not applicable for all solid-state reactions, as the $\tilde{\mu}_j \approx \mu_j$ approximation does not hold when species exhibit variable oxidation states, or there exists a significant amount of charge transfer between species, both of which are known to occur in many solid-state reactions involving transition metals. In the present study, all phases considered have species in only a single oxidation state: +2 for Ba, +4 for Ti and -2 for O. Furthermore, the phenomenological expression for the flux (Equation \ref{eqn: fluxeqn}) is defined under the center-of-mass frame of reference, which imposes constraints on the total number of independent material fluxes in the system for incompressible systems. These constraints are described in the S.I.

\subsection{Simulation of solid-state synthesis reactions with ReactCA}
\label{sec:rxn-ca-description}
A cellular automaton framework was recently developed (ReactCA)\cite{gallant2024cellular} to simulate the evolution of mixtures of powders over the course of a solid-state reactions. The reaction vessel is discretized into voxels (cells) and each voxel represents a  powder particle of a given phase. At a given temperature, all possible reactions that can occur between any other powders (phases) in the system are enumerated combinatorially with an approach developed as part of a previous work \cite{mcdermott_graph-based_2021}, and are scored using a scoring function. The system is initialized with a random distribution of two phases (in this case representing two well mixed precursor powders of uniform size). At every step of the simulation, two reactants are selected, and a reaction is probabilistically selected from the enumerated reactions based on their scores, and the reactants are then probabilistically replaced with products based on the stoichiometry of the reaction. This process is repeated millions of times throughout the simulation to accurately model the evolution of phases within the reaction vessel. In the original publication \cite{gallant2024cellular}, reaction scores (i.e., rates) between neighboring particles were estimated based on the reaction thermodynamics (as estimated by zero-temperature formation energies from the Materials Project in conjunction with a machine learning estimate of the vibrational Gibbs free energy \cite{bartel_physical_2018}) and machine learning estimates of the melting point of the precursor materials \cite{hong_melting_2022} (a preliminary proxy for kinetic facility of the reaction). In this work, given a reaction $\alpha + \beta \rightarrow \gamma$, we incorporate the diffusive fluxes using a new scoring function $S$ for ReactCA, as: 

\begin{equation}
\label{eqn:new_score}
\begin{split}
    & S = \sigma_1(\frac{K_{D}}{r_0^2s} * \frac{\Delta G^*}{k_BT}) * \sigma_2 (\frac{T}{T_{m, reactant}}) \\
    & \sigma_1(x) = \frac{1}{3}\log(1+\exp(ax)) \\
    & \sigma_2(x) = \frac{1}{2}\log(1+\exp(bx - c)) \\
    & \Delta G^* = 1 + \text{erf}(-d(\Delta G_{rxn} + e)) \\
    & K_D = \Sigma_i \Sigma_j \frac{|L^{\gamma}_{ij} V^{\gamma}\cdot \text{min}(\mu_j^\alpha - \mu_j^\beta)|}{n^{\gamma}_i}
\end{split}
\end{equation}

The form of this scoring function is an extension to the scorer present in ReactCA \cite{gallant2024cellular}, employing the same constants $a, b, c, d, e$. The factor $s$ is set equal for all reactions considered, to appropriately scale the scores to be stable in the CA simulation and to ensure the reaction between the precursors initiate at temperatures  above the decomposition temperature of \ce{BaCO3} ($s=10^{3}$). This scoring function has a ``soft'' activation towards Tammann's rule \cite{merkle2005tammann}, implemented through the $\sigma_2$ part of the function, and a ``switching-on'' behavior towards reactions with negative reaction free energies ($\Delta G_{rxn}$) through the error function. For more details on these aspects of the scoring function, we refer to Ref~ \cite{gallant2024cellular}. In this work, we implement explicit, quantitative transport diffusion constants in the amorphous product phase ($K_D$) through a soft activation with the $\sigma_1$ function, which is also a soft-plus function. This function also takes the modified $\Delta G_{rxn}$ and the temperature into account. Hence, below the Tammann temperature ($\frac{T}{T_{m, reactant}} \leq 0.67$), the scores are non-zero, but only slightly positive. Above the Tammann temperature, reactions with a slightly positive $\Delta G_{rxn}$ or with low ionic fluxes have a non-zero but small score assigned to them. With increasing reaction temperature, the $\sigma_2$ part of the function increases, thereby elevating the reaction rate. Similarly, raising the temperature also increases the diffusion rate constant $K_D$ (often in a super-linear fashion), which generally boosts the output of the $\sigma_1$ part of the scoring function. 
However, as demonstrated in this work, $K_D$  approaches a saturation point with increasing temperature, which in turn causes the output of $\sigma_1$  to also level off or even decrease at very high temperatures. This effect leads to the thermodynamic regime of phase selectivity when the reaction temperatures are close to or above the melting point. Finally, the scoring function maintains backward compatibility with the old scoring function, which is particularly important when kinetic transport parameters are not available for some or all phases in the system. 

\subsection{Density Functional Theory Calculations}
\label{sec:dft}
DFT single-point calculations and ab-initio molecular dynamics (AIMD) calculations were performed using the Vienna Ab-Initio Simulation Package (VASP) \cite{kresse1996vasp4, kresse1996vasp3}, the Perdew-Burke-Ernzerhof \cite{perdew1996generalized} formulation of generalized gradient approximation with projector-augmented wave potentials \cite{blochl1994projector, kresse1999ultrasoft}. Furthermore, to minimize computation requirements, all AIMD calculations were performed using the $\Gamma$-point only and were non-spin polarized. For more details on the parameters used, we refer to the \texttt{MPMorphMDSet} class in \texttt{atomate2} \cite{ganose_atomate2_2024}. To construct high temperature phase diagrams, a machine-learned estimator for the vibrational contribution to entropy \cite{bartel_physical_2018} was used to estimate the finite temperature vibrational Gibbs free energies for all phases. 

\section{Acknowledgements}

This work was intellectually led by D2S2, the core program KCD2S2, which is supported by the U.S. Department of Energy, Office of Science, Office of Basic Energy Sciences, Materials Sciences and Engineering Division under contract no. DE-AC02-05-CH11231.  This research used resources of the National Energy Research Scientific Computing Center (NERSC), a U.S. Department of Energy Office of Science User Facility operated under Contract No. DE-AC02-05CH11231.

\section{Author Contributions}
\textbf{V.K}: Conceptualization, Methodology, Software, Investigation, Writing - Original Draft, Writing - Review \& Editing. \textbf{M.G}: Conceptualization, Methodology, Software, Writing - Review \& Editing. \textbf{Y.F}: Validation. \textbf{G.C}: Supervision, Project administration. \textbf{K.A.P}: Conceptualization, Writing - Review \& Editing, Supervision, Project administration.

\section{Competing Interests}

All authors declare no financial or non-financial competing interests.

\printbibliography

@article{disalvo_solid-state_1990,
	title = {Solid-{State} {Chemistry}: a {Rediscovered} {Chemical} {Frontier}},
	volume = {247},
	issn = {0036-8075, 1095-9203},
	shorttitle = {Solid-{State} {Chemistry}},
	doi = {10.1126/science.247.4943.649},
	language = {en},
	number = {4943},
	journal = {Science},
	author = {DiSalvo, Francis J.},
	year = {1990},
	pages = {649--655},
}

@book{schmalzried2008chemical,
  title={Chemical kinetics of solids},
  author={Schmalzried, Hermann},
  year={2008},
  publisher={John Wiley \& Sons}
}

@article{jain_commentary_2013,
	title = {Commentary: {The} {Materials} {Project}: {A} materials genome approach to accelerating materials innovation},
	volume = {1},
	issn = {2166-532X},
	shorttitle = {Commentary},
	url = {https://doi.org/10.1063/1.4812323},
	doi = {10.1063/1.4812323},
	number = {1},
	urldate = {2024-03-15},
	journal = {APL Materials},
	author = {Jain, Anubhav and Ong, Shyue Ping and Hautier, Geoffroy and Chen, Wei and Richards, William Davidson and Dacek, Stephen and Cholia, Shreyas and Gunter, Dan and Skinner, David and Ceder, Gerbrand and Persson, Kristin A.},
	month = jul,
	year = {2013},
	pages = {011002},
}

@article{mcdermott_graph-based_2021,
	title = {A graph-based network for predicting chemical reaction pathways in solid-state materials synthesis},
	volume = {12},
	issn = {2041-1723},
	url = {https://www.nature.com/articles/s41467-021-23339-x},
	doi = {10.1038/s41467-021-23339-x},
	language = {en},
	number = {1},
	urldate = {2023-09-01},
	journal = {Nature Communications},
	author = {McDermott, Matthew J. and Dwaraknath, Shyam S. and Persson, Kristin A.},
	month = may,
	year = {2021},
	pages = {3097},
}

@article{szymanski_autonomous_2023,
	title = {Autonomous and dynamic precursor selection for solid-state materials synthesis},
	volume = {14},
	copyright = {2023 The Author(s)},
	issn = {2041-1723},
	url = {https://www.nature.com/articles/s41467-023-42329-9},
	doi = {10.1038/s41467-023-42329-9},
	language = {en},
	number = {1},
	urldate = {2024-03-14},
	journal = {Nature Communications},
	author = {Szymanski, Nathan J. and Nevatia, Pragnay and Bartel, Christopher J. and Zeng, Yan and Ceder, Gerbrand},
	month = oct,
	year = {2023},
	note = {Publisher: Nature Publishing Group},
	pages = {6956},
}

@article{
szymanski_quantifying_2024,
author = {Nathan J. Szymanski  and Young-Woon Byeon  and Yingzhi Sun  and Yan Zeng  and Jianming Bai  and Martin Kunz  and Dong-Min Kim  and Brett A. Helms  and Christopher J. Bartel  and Haegyeom Kim  and Gerbrand Ceder },
title = {Quantifying the regime of thermodynamic control for solid-state reactions during ternary metal oxide synthesis},
journal = {Science Advances},
volume = {10},
number = {27},
pages = {eadp3309},
year = {2024},
doi = {10.1126/sciadv.adp3309},
URL = {https://www.science.org/doi/abs/10.1126/sciadv.adp3309},
eprint = {https://www.science.org/doi/pdf/10.1126/sciadv.adp3309}}

@article{khawam2006solid,
  title={Solid-state kinetic models: basics and mathematical fundamentals},
  author={Khawam, Ammar and Flanagan, Douglas R},
  journal={The journal of physical chemistry B},
  volume={110},
  number={35},
  pages={17315--17328},
  year={2006},
  publisher={ACS Publications}
}

@article{frade1992reexamination,
  title={Reexamination of the Basic Theoretical Model for the Kinetics of Solid--State Reactions},
  author={Frade, Jorge R and Cable, Michael},
  journal={Journal of the American Ceramic Society},
  volume={75},
  number={7},
  pages={1949--1957},
  year={1992},
  publisher={Wiley Online Library}
}

@article{cordova_synthesis_2020,
author = {Cordova, Dmitri Leo M. and Johnson, David C.},
title = {Synthesis of Metastable Inorganic Solids with Extended Structures},
journal = {ChemPhysChem},
volume = {21},
number = {13},
pages = {1345-1368},
doi = {https://doi.org/10.1002/cphc.202000199},
url = {https://chemistry-europe.onlinelibrary.wiley.com/doi/abs/10.1002/cphc.202000199},
eprint = {https://chemistry-europe.onlinelibrary.wiley.com/doi/pdf/10.1002/cphc.202000199},
year = {2020}
}

@article{Beauger1983SynthesisRO,
  title={Synthesis reaction of metatitanate BaTiO3},
  author={Alain Beauger and J. C. Mutin and J. C. Ni{\`e}pce},
  journal={Journal of Materials Science},
  year={1983},
  volume={18},
  pages={3543-3550},
  url={https://api.semanticscholar.org/CorpusID:93995149}
}

@article{brzozowski2002baco,
  title={BaCO 3--TiO 2 solid state reaction: a kinetic study},
  author={Brzozowski, E and Sanchez, J and Castro, MS},
  journal={Journal of Materials Synthesis and Processing},
  volume={10},
  pages={1--5},
  year={2002},
  publisher={Springer}
}

@techreport{osti_6480472,
  author       = {Bierach, J W},
  title        = {The formation of barium titanate ceramics by solid state reaction},
  institution  = {Lawrence Berkeley Lab., CA (USA)},
  url          = {https://www.osti.gov/biblio/6480472},
  place        = {United States},
  year         = {1988},
  month        = {05}}

@article{mcdermott2023assessing,
  title={Assessing thermodynamic selectivity of solid-state reactions for the predictive synthesis of inorganic materials},
  author={McDermott, Matthew J and McBride, Brennan C and Regier, Corlyn E and Tran, Gia Thinh and Chen, Yu and Corrao, Adam A and Gallant, Max C and Kamm, Gabrielle E and Bartel, Christopher J and Chapman, Karena W and others},
  journal={ACS Central Science},
  volume={9},
  number={10},
  pages={1957--1975},
  year={2023},
  publisher={ACS Publications}
}

@article{bartel_physical_2018,
	title = {Physical descriptor for the {Gibbs} energy of inorganic crystalline solids and temperature-dependent materials chemistry},
	volume = {9},
	issn = {2041-1723},
	url = {https://www.nature.com/articles/s41467-018-06682-4},
	doi = {10.1038/s41467-018-06682-4},
	language = {en},
	number = {1},
	urldate = {2024-04-26},
	journal = {Nature Communications},
	author = {Bartel, Christopher J. and Millican, Samantha L. and Deml, Ann M. and Rumptz, John R. and Tumas, William and Weimer, Alan W. and Lany, Stephan and Stevanović, Vladan and Musgrave, Charles B. and Holder, Aaron M.},
	month = oct,
	year = {2018},
	pages = {4168},
}

@article{buscaglia2008solid,
  title={Solid-state synthesis of nanocrystalline BaTiO3: reaction kinetics and powder properties},
  author={Buscaglia, Maria Teresa and Bassoli, Marta and Buscaglia, Vincenzo and Vormberg, Reinhard},
  journal={Journal of the American Ceramic Society},
  volume={91},
  number={9},
  pages={2862--2869},
  year={2008},
  publisher={Wiley Online Library}
}

@article{wu1988factors,
  title={Factors affecting the formation of Ba2Ti9O20},
  author={WU, JENN-MING and WANG, HONG-WEN},
  journal={Journal of the American Ceramic Society},
  volume={71},
  number={10},
  pages={869--875},
  year={1988},
  publisher={Wiley Online Library}
}

@article{gallant2024cellular,
  title={A Cellular Automaton Simulation for Predicting Phase Evolution in Solid-State Reactions},
  author={Gallant, Max C and McDermott, Matthew J and Li, Bryant and Persson, Kristin A},
  journal={Chemistry of Materials},
  year={2024},
  publisher={ACS Publications}
}

@article{merkle2005tammann,
  title={On the tammann--rule},
  author={Merkle, Rotraut and Maier, Joachim},
  journal={Zeitschrift f{\"u}r anorganische und allgemeine Chemie},
  volume={631},
  number={6-7},
  pages={1163--1166},
  year={2005},
  publisher={Wiley Online Library}
}

@article{bondioli_kinetic_1998,
	title = {Kinetic {Study} of {Conventional} {Solid}-{State} {Synthesis} of {BaTiO}$_{\textrm{3}}$ by \textit{in situ} {HT}-{XRD}},
	volume = {278-281},
	issn = {1662-9752},
	url = {https://www.scientific.net/MSF.278-281.379},
	doi = {10.4028/www.scientific.net/MSF.278-281.379},
	language = {en},
	urldate = {2023-10-04},
	journal = {Materials Science Forum},
	author = {Bondioli, Federica and Bonamartini Corradi, A. and Ferrari, Anna Maria and Manfredini, Tiziano and Pellacani, Gian Carlo},
	month = apr,
	year = {1998},
	pages = {379--383},
}

@article{o1975preparation,
  title={Preparation of BaTi5O11 by Solid-State Reaction},
  author={O'BRYAN JR, HENRY M and THOMSON JR, JOHN},
  journal={Journal of the American Ceramic Society},
  volume={58},
  number={9-10},
  pages={454--454},
  year={1975},
  publisher={Wiley Online Library}
}

@article{obryan1983ba2ti9o20,
author = {O'BRYAN, HENRY M. and THOMSON, JOHN},
title = {Ba2Ti9O20 Phase Equilibria},
journal = {Journal of the American Ceramic Society},
volume = {66},
number = {1},
pages = {66-68},
doi = {https://doi.org/10.1111/j.1151-2916.1983.tb09970.x},
url = {https://ceramics.onlinelibrary.wiley.com/doi/abs/10.1111/j.1151-2916.1983.tb09970.x},
eprint = {https://ceramics.onlinelibrary.wiley.com/doi/pdf/10.1111/j.1151-2916.1983.tb09970.x},
year = {1983}
}

@article{zhu2010formation,
  title={Formation and stability of ferroelectric BaTi2O5},
  author={Zhu, Na and West, Anthony R},
  journal={Journal of the American Ceramic Society},
  volume={93},
  number={1},
  pages={295--300},
  year={2010},
  publisher={Wiley Online Library}
}

@article{cheng2022materials,
  title={Materials design principles of amorphous cathode coatings for lithium-ion battery applications},
  author={Cheng, Jianli and Fong, Kara D and Persson, Kristin A},
  journal={Journal of Materials Chemistry A},
  volume={10},
  number={41},
  pages={22245--22256},
  year={2022},
  publisher={Royal Society of Chemistry}
}

@article{zimmermann2020local,
  title={Local structure order parameters and site fingerprints for quantification of coordination environment and crystal structure similarity},
  author={Zimmermann, Nils ER and Jain, Anubhav},
  journal={RSC advances},
  volume={10},
  number={10},
  pages={6063--6081},
  year={2020},
  publisher={Royal Society of Chemistry}
}

@article{ong2013python,
  title={Python Materials Genomics (pymatgen): A robust, open-source python library for materials analysis},
  author={Ong, Shyue Ping and Richards, William Davidson and Jain, Anubhav and Hautier, Geoffroy and Kocher, Michael and Cholia, Shreyas and Gunter, Dan and Chevrier, Vincent L and Persson, Kristin A and Ceder, Gerbrand},
  journal={Computational Materials Science},
  volume={68},
  pages={314--319},
  year={2013},
  publisher={Elsevier}
}

@article{neilson_modernist_2023,
	title = {Modernist {Materials} {Synthesis}: {Finding} {Thermodynamic} {Shortcuts} with {Hyperdimensional} {Chemistry}},
	volume = {38},
	issn = {0884-2914, 2044-5326},
	shorttitle = {Modernist {Materials} {Synthesis}},
	url = {http://arxiv.org/abs/2303.11915},
	doi = {10.1557/s43578-023-01037-2},
	language = {en},
	number = {11},
	urldate = {2023-09-12},
	journal = {Journal of Materials Research},
	author = {Neilson, James R. and McDermott, Matthew J. and Persson, Kristin A.},
	month = jun,
	year = {2023},
	note = {arXiv:2303.11915 [cond-mat]},
	pages = {2885--2893},
}

@article{aykol_rational_2021,
	title = {Rational {Solid}-{State} {Synthesis} {Routes} for {Inorganic} {Materials}},
	volume = {143},
	issn = {0002-7863, 1520-5126},
	url = {https://pubs.acs.org/doi/10.1021/jacs.1c04888},
	doi = {10.1021/jacs.1c04888},
	language = {en},
	number = {24},
	urldate = {2023-09-06},
	journal = {Journal of the American Chemical Society},
	author = {Aykol, Muratahan and Montoya, Joseph H. and Hummelshøj, Jens},
	month = jun,
	year = {2021},
	pages = {9244--9259},
}

@article{lee_2007_phasediagram,
author = {Lee, Soonil and Randall, Clive A. and Liu, Zi-Kui},
title = {Modified Phase Diagram for the Barium Oxide–Titanium Dioxide System for the Ferroelectric Barium Titanate},
journal = {Journal of the American Ceramic Society},
volume = {90},
number = {8},
pages = {2589-2594},
doi = {https://doi.org/10.1111/j.1551-2916.2007.01794.x},
url = {https://ceramics.onlinelibrary.wiley.com/doi/abs/10.1111/j.1551-2916.2007.01794.x},
eprint = {https://ceramics.onlinelibrary.wiley.com/doi/pdf/10.1111/j.1551-2916.2007.01794.x},
year = {2007}
}

@article{fong_transport_2020,
  title={Transport phenomena in electrolyte solutions: Nonequilibrium thermodynamics and statistical mechanics},
  author={Fong, Kara D and Bergstrom, Helen K and McCloskey, Bryan D and Mandadapu, Kranthi K},
  journal={AIChE Journal},
  volume={66},
  number={12},
  pages={e17091},
  year={2020},
  publisher={Wiley Online Library}
}

@article{fong2020onsager,
  title={Onsager transport coefficients and transference numbers in polyelectrolyte solutions and polymerized ionic liquids},
  author={Fong, Kara D and Self, Julian and McCloskey, Bryan D and Persson, Kristin A},
  journal={Macromolecules},
  volume={53},
  number={21},
  pages={9503--9512},
  year={2020},
  publisher={ACS Publications}
}

@article{drautz2019atomic,
  title={Atomic cluster expansion for accurate and transferable interatomic potentials},
  author={Drautz, Ralf},
  journal={Physical Review B},
  volume={99},
  number={1},
  pages={014104},
  year={2019},
  publisher={APS}
}

@article{aykol_thermodynamic_2018,
	title = {Thermodynamic limit for synthesis of metastable inorganic materials},
	volume = {4},
	url = {https://www.science.org/doi/10.1126/sciadv.aaq0148},
	doi = {10.1126/sciadv.aaq0148},
	number = {4},
	urldate = {2023-09-08},
	journal = {Science Advances},
	author = {Aykol, Muratahan and Dwaraknath, Shyam S. and Sun, Wenhao and Persson, Kristin A.},
	month = apr,
	year = {2018},
	note = {Publisher: American Association for the Advancement of Science},
	pages = {eaaq0148},
}

@article{zheng2024ab,
  title={The ab initio non-crystalline structure database: empowering machine learning to decode diffusivity},
  author={Zheng, Hui and Sivonxay, Eric and Christensen, Rasmus and Gallant, Max and Luo, Ziyao and McDermott, Matthew and Huck, Patrick and Smedskj{\ae}r, Morten M and Persson, Kristin A},
  journal={npj Computational Materials},
  volume={10},
  number={1},
  pages={295},
  year={2024},
  publisher={Nature Publishing Group UK London}
}

@software{ganose_atomate2_2024,
  author = {Ganose, Alex and Riebesell, Janosh and George, J. and Shen, Jimmy and S. Rosen, Andrew and Ashok Naik, Aakash and nwinner and Wen, Mingjian and rdguha1995 and Kuner, Matthew and Petretto, Guido and Zhu, Zhuoying and Horton, Matthew and Sahasrabuddhe, Hrushikesh and Kaplan, Aaron and Schmidt, Jonathan and Ertural, Christina and Kingsbury, Ryan and McDermott, Matt and Goodall, Rhys and Bonkowski, Alexander and Purcell, Thomas and Zügner, Daniel and Qi, Ji},
  doi = {10.5281/zenodo.10677081},
  license = {cc-by-4.0},
  month = jan,
  title = {atomate2},
  url = {https://github.com/materialsproject/atomate2},
  version = {0.0.13},
  year = {2024}
}

@article{packmol,
author = {Martínez, L. and Andrade, R. and Birgin, E. G. and Martínez, J. M.},
title = {PACKMOL: A package for building initial configurations for molecular dynamics simulations},
journal = {Journal of Computational Chemistry},
volume = {30},
number = {13},
pages = {2157-2164},
doi = {https://doi.org/10.1002/jcc.21224},
url = {https://onlinelibrary.wiley.com/doi/abs/10.1002/jcc.21224},
eprint = {https://onlinelibrary.wiley.com/doi/pdf/10.1002/jcc.21224},
year = {2009}
}

@article{rosen2024jobflow,
  title={Jobflow: Computational workflows made simple},
  author={Rosen, Andrew S and Gallant, Max and George, Janine and Riebesell, Janosh and Sahasrabuddhe, Hrushikesh and Shen, Jimmy-Xuan and Wen, Mingjian and Evans, Matthew L and Petretto, Guido and Waroquiers, David and others},
  journal={Journal of Open Source Software},
  volume={9},
  number={93},
  pages={5995},
  year={2024}
}

@article{jain2015fireworks,
  title={FireWorks: a dynamic workflow system designed for high-throughput applications},
  author={Jain, Anubhav and Ong, Shyue Ping and Chen, Wei and Medasani, Bharat and Qu, Xiaohui and Kocher, Michael and Brafman, Miriam and Petretto, Guido and Rignanese, Gian-Marco and Hautier, Geoffroy and others},
  journal={Concurrency and Computation: Practice and Experience},
  volume={27},
  number={17},
  pages={5037--5059},
  year={2015},
  publisher={Wiley Online Library}
}

@article{lysogorskiy2021performant,
  title={Performant implementation of the atomic cluster expansion (PACE) and application to copper and silicon},
  author={Lysogorskiy, Yury and Oord, Cas van der and Bochkarev, Anton and Menon, Sarath and Rinaldi, Matteo and Hammerschmidt, Thomas and Mrovec, Matous and Thompson, Aidan and Cs{\'a}nyi, G{\'a}bor and Ortner, Christoph and others},
  journal={npj computational materials},
  volume={7},
  number={1},
  pages={97},
  year={2021},
  publisher={Nature Publishing Group UK London}
}

@article{bochkarev2022parameterization,
  title = {Efficient parametrization of the atomic cluster expansion},
  author = {Bochkarev, Anton and Lysogorskiy, Yury and Menon, Sarath and Qamar, Minaam and Mrovec, Matous and Drautz, Ralf},
  journal = {Phys. Rev. Mater.},
  volume = {6},
  issue = {1},
  pages = {013804},
  numpages = {18},
  year = {2022},
  month = {Jan},
  publisher = {American Physical Society},
  doi = {10.1103/PhysRevMaterials.6.013804},
  url = {https://link.aps.org/doi/10.1103/PhysRevMaterials.6.013804}
}

@article{erhard2024modelling,
  title={Modelling atomic and nanoscale structure in the silicon--oxygen system through active machine learning},
  author={Erhard, Linus C and Rohrer, Jochen and Albe, Karsten and Deringer, Volker L},
  journal={Nature Communications},
  volume={15},
  number={1},
  pages={1927},
  year={2024},
  publisher={Nature Publishing Group UK London}
}

@article{lysogorskiy2023active,
  title={Active learning strategies for atomic cluster expansion models},
  author={Lysogorskiy, Yury and Bochkarev, Anton and Mrovec, Matous and Drautz, Ralf},
  journal={Physical Review Materials},
  volume={7},
  number={4},
  pages={043801},
  year={2023},
  publisher={APS}
}

@article{schmalzried_build-up_1986,
	title = {The {Build}-{Up} of {Internal} {Pressures} {During} {Compound} {Formation}},
	volume = {148},
	issn = {2196-7156, 0942-9352},
	url = {https://www.degruyter.com/document/doi/10.1524/zpch.1986.148.1.021/html},
	doi = {10.1524/zpch.1986.148.1.021},
	language = {en},
	number = {1},
	urldate = {2023-11-16},
	journal = {Zeitschrift für Physikalische Chemie},
	author = {Schmalzried, H. and Pfeiffer, Th.},
	month = jan,
	year = {1986},
	pages = {21--32},
}

@article{lu1998kinetic,
  title={Kinetic analysis of the serial reactions of lead magnesium tungstate ceramics using a multiple core-shell model},
  author={Lu, Chung-Hsin and Lee, Jiun-Ting},
  journal={Journal of materials science},
  volume={33},
  pages={2121--2127},
  year={1998},
  publisher={Springer}
}

@article{dheurle_theoretical_1995,
  title={Theoretical considerations about phase growth and phase formation},
  author={d’Heurle, FM},
  journal={MRS Online Proceedings Library},
  volume={402},
  number={1},
  pages={3--14},
  year={1995},
  publisher={Springer}
}

@article{hong_melting_2022,
  title={Melting temperature prediction using a graph neural network model: From ancient minerals to new materials},
  author={Hong, Qi-Jun and Ushakov, Sergey V and van de Walle, Axel and Navrotsky, Alexandra},
  journal={Proceedings of the National Academy of Sciences},
  volume={119},
  number={36},
  pages={e2209630119},
  year={2022},
  publisher={National Acad Sciences}
}

@article{kresse1996vasp4,
	title = {Efficiency of ab-initio total energy calculations for metals and semiconductors using a plane-wave basis set},
	volume = {6},
	issn = {09270256},
	url = {https://linkinghub.elsevier.com/retrieve/pii/0927025696000080},
	doi = {10.1016/0927-0256(96)00008-0},
	language = {en},
	number = {1},
	urldate = {2021-04-29},
	journal = {Computational Materials Science},
	author = {Kresse, G. and Furthmüller, J.},
	month = jul,
	year = {1996},
	pages = {15--50},
}

@article{kresse1996vasp3,
	title = {Efficient iterative schemes for \textit{ab initio} total-energy calculations using a plane-wave basis set},
	volume = {54},
	issn = {0163-1829, 1095-3795},
	url = {https://link.aps.org/doi/10.1103/PhysRevB.54.11169},
	doi = {10.1103/PhysRevB.54.11169},
	language = {en},
	number = {16},
	urldate = {2021-04-29},
	journal = {Physical Review B},
	author = {Kresse, G. and Furthmüller, J.},
	month = oct,
	year = {1996},
	pages = {11169--11186},
}

@article{perdew1996generalized,
  title = {Generalized Gradient Approximation Made Simple},
  author = {Perdew, John P. and Burke, Kieron and Ernzerhof, Matthias},
  journal = {Phys. Rev. Lett.},
  volume = {77},
  issue = {18},
  pages = {3865--3868},
  numpages = {0},
  year = {1996},
  month = {Oct},
  publisher = {American Physical Society},
  doi = {10.1103/PhysRevLett.77.3865},
  url = {https://link.aps.org/doi/10.1103/PhysRevLett.77.3865}
}

@article{blochl1994projector,
  title = {Projector augmented-wave method},
  author = {Bl\"ochl, P. E.},
  journal = {Phys. Rev. B},
  volume = {50},
  issue = {24},
  pages = {17953--17979},
  numpages = {0},
  year = {1994},
  month = {Dec},
  publisher = {American Physical Society},
  doi = {10.1103/PhysRevB.50.17953},
  url = {https://link.aps.org/doi/10.1103/PhysRevB.50.17953}
}

@article{kresse1999ultrasoft,
  title = {From ultrasoft pseudopotentials to the projector augmented-wave method},
  author = {Kresse, G. and Joubert, D.},
  journal = {Phys. Rev. B},
  volume = {59},
  issue = {3},
  pages = {1758--1775},
  numpages = {0},
  year = {1999},
  month = {Jan},
  publisher = {American Physical Society},
  doi = {10.1103/PhysRevB.59.1758},
  url = {https://link.aps.org/doi/10.1103/PhysRevB.59.1758}
}

@article{yokokawa1999generalized,
  title={Generalized chemical potential diagram and its applications to chemical reactions at interfaces between dissimilar materials},
  author={Yokokawa, H},
  journal={Journal of phase equilibria},
  volume={20},
  number={3},
  pages={258--287},
  year={1999},
  publisher={Springer}
}

\newpage

\clearpage

\setcounter{page}{1}
\setcounter{section}{0}
\setcounter{table}{0}
\setcounter{figure}{0}
\setcounter{equation}{1}

\renewcommand{\thepage}{S\arabic{page}}
\renewcommand{\thesection}{S\arabic{section}}
\renewcommand{\theequation}{S\arabic{equation}}
\renewcommand{\thetable}{S\arabic{table}}
\renewcommand{\thefigure}{S\arabic{figure}}

\clearpage

\section*{\centering Supplementary Information}

\section{Validation of trained Atomic Cluster Expansion Potential}

We reiterate that the purpose of the trained potential is to perform high-temperature, long time-scale molecular dynamics of amorphous configurations in the Ba-Ti-O chemical space. As such, we here demonstrate that the trained potential is accurate for this particular task. 

\subsection{Training and Test Loss}

\begin{figure}[htbp]
    \centering
    \includegraphics[width=\linewidth]{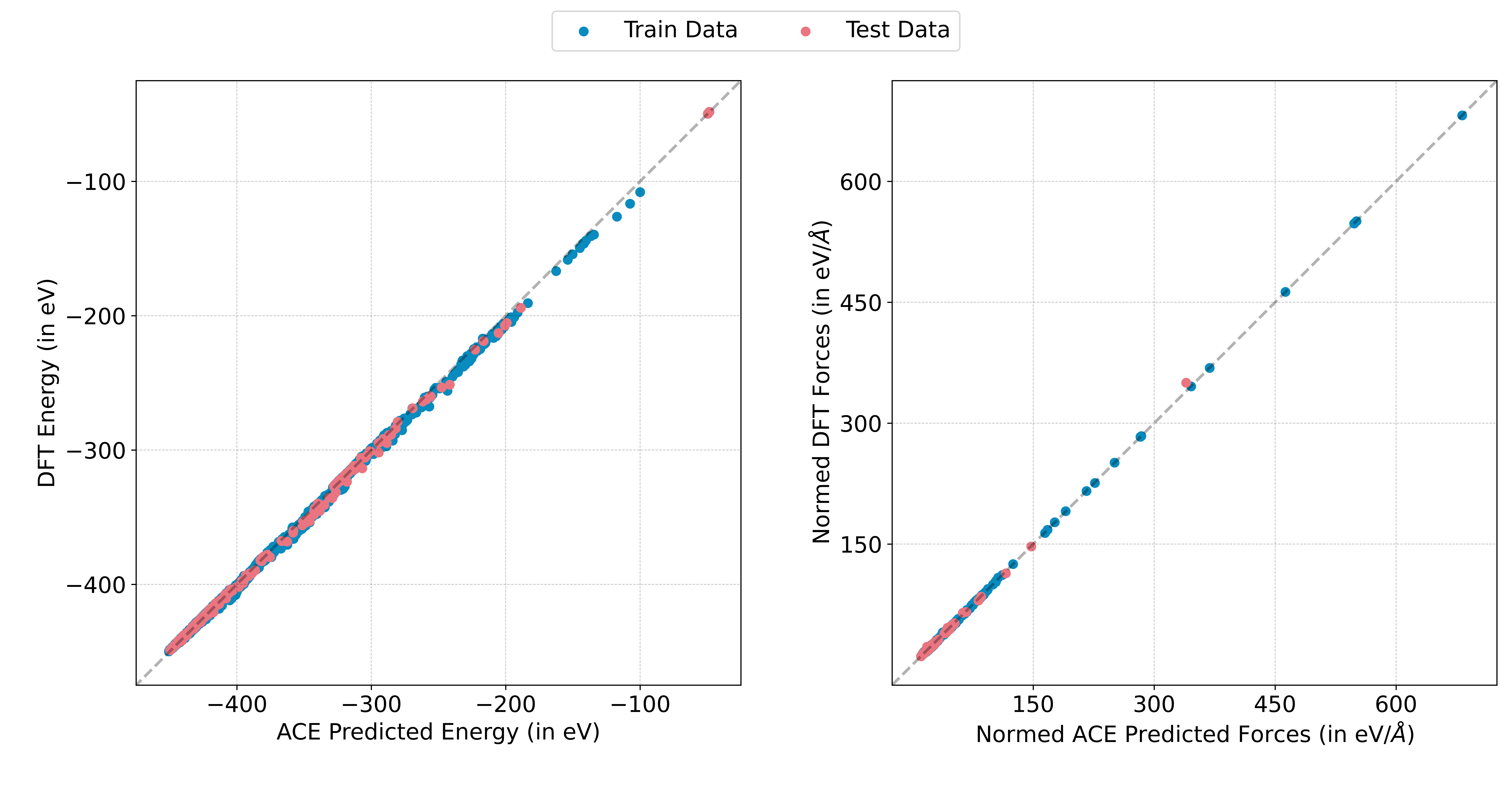}
    \caption{Parity plot of predicted (ACE) and ground truth for Ba-Ti-O system (DFT), A: energies, B: norm of forces.}
    \label{fig:traintest}
\end{figure}

The RMSE of energies and forces of the trained potential for the Ba-Ti-O system are given in Table \ref{tab:traintesterrors}. A parity plot comparing the predictions for energies and norm of forces of the potential to the ground truth data is shown in Figure \ref{fig:traintest}. 
\begin{table}[htbp]
    \centering
    \begin{tabular}{ c | c | c }
        Metric & Energy (in meV/atom) & Force (in meV/A) \\
        \hline
        RMSE & 16.50 & 239.20 \\
        MAE & 10.90 & 168.31
    \end{tabular}
    \caption{Metrics to quantify performance of the trained ACE potential w.r.t ground truth DFT data. RMSE: Root Mean Squared Error, MAE: Mean Absolute Error.}
    \label{tab:traintesterrors}
\end{table}

\subsection{Binary elemental interactions}

\begin{figure}[htbp]
    \centering
    \includegraphics[width=0.5\linewidth]{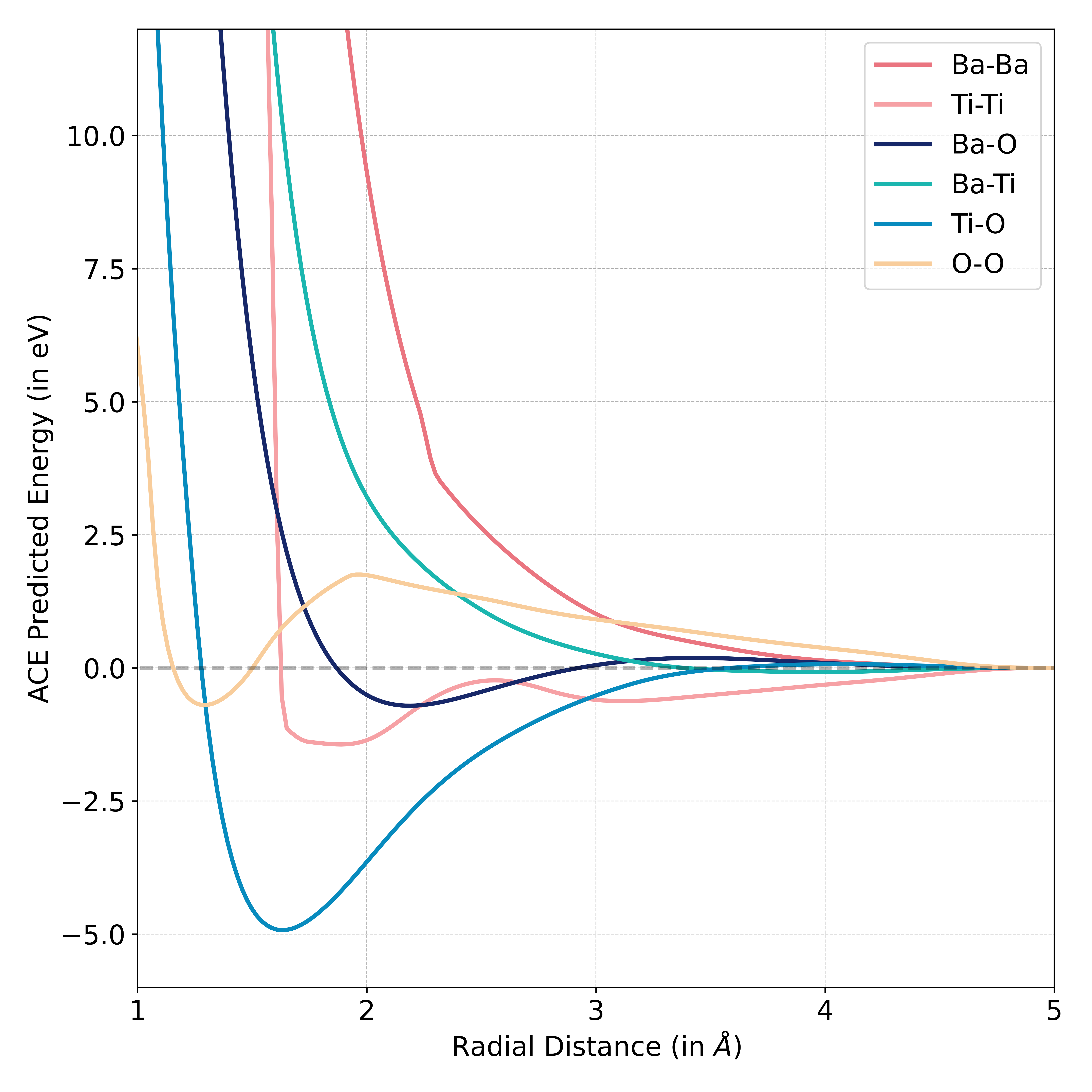}
    \caption{Energy vs radial distance plot for different dimer configurations, computed using the trained ACE potential.}
    \label{fig:binary}
\end{figure}

A quick way to assess the quality of an interatomic potential is to verify whether it accurately captures the equilibrium bond lengths, bond energies, and core repulsive behavior observed in physical bonds. This can be done by evaluating the energy of a dimer in vacuum, systematically varying the atomic species and the radial distance between the two atoms. The result is shown in Figure \ref{fig:binary}, which demonstrates that the model captures the behavior expected of chemical bonds, showing clear repulsion at short distances, while also displaying approximately correct bond lengths for the Ti-O and Ba-O dimers.

\subsection{Equation of State test}

\begin{figure}[htbp]
    \centering
    \includegraphics[width=0.95\linewidth]{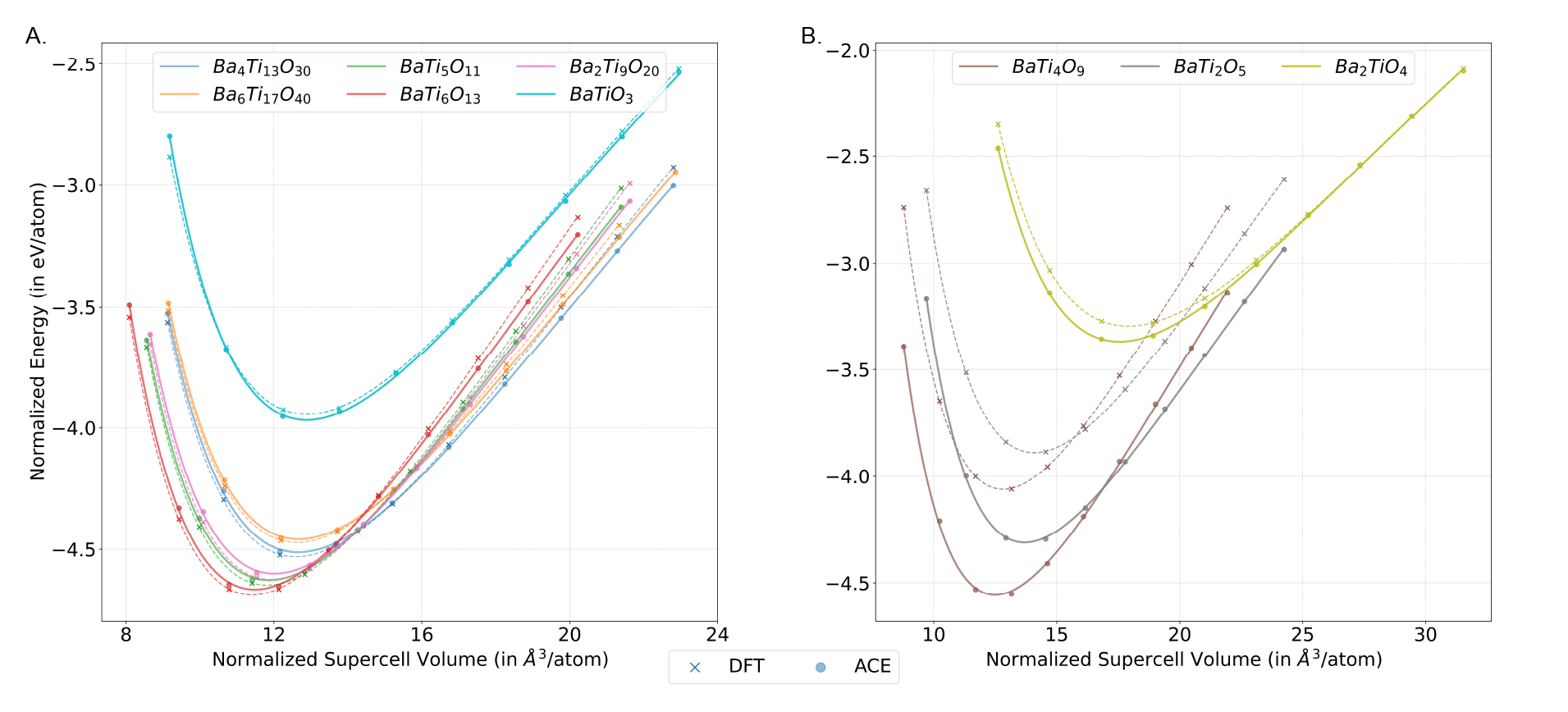}
    \caption{Energy vs Volume plots for crystalline structures taken from MP in the Ba-Ti-O system.}
    \label{fig:bto_eos}
\end{figure}

To measure the efficacy of the potential for bulk systems, structures corresponding to experimentally observed phases are taken from the Materials Project, compressed and stretched by 30\% to emulate the volume change. The energy for each volume is computed using both the trained ACE potential and single-point DFT in VASP. A Birch-Murnaghan equation of state is fit, as shown in Figures \ref{fig:bto_eos}. The potential displays the energy-volume well that is expected from stable crystals for all structures considered, matching well the DFT energies for most structures. Remarkably, even for structures with significant deviations in the EOS, shown in Figure \ref{fig:bto_eos}B, the computed equilibrium volumes and bulk modulus are within 5\% of those computed from DFT data. This implies that the potential has captured the curvature of the potential energy landscape well for all compositions of interest in this system, rendering it suitable for MD simulations which rely on the predicted forces rather than energies. 

\subsection{RDFs and short range order}

\begin{figure}[h!]
    \centering
    \includegraphics[width=0.95\linewidth]{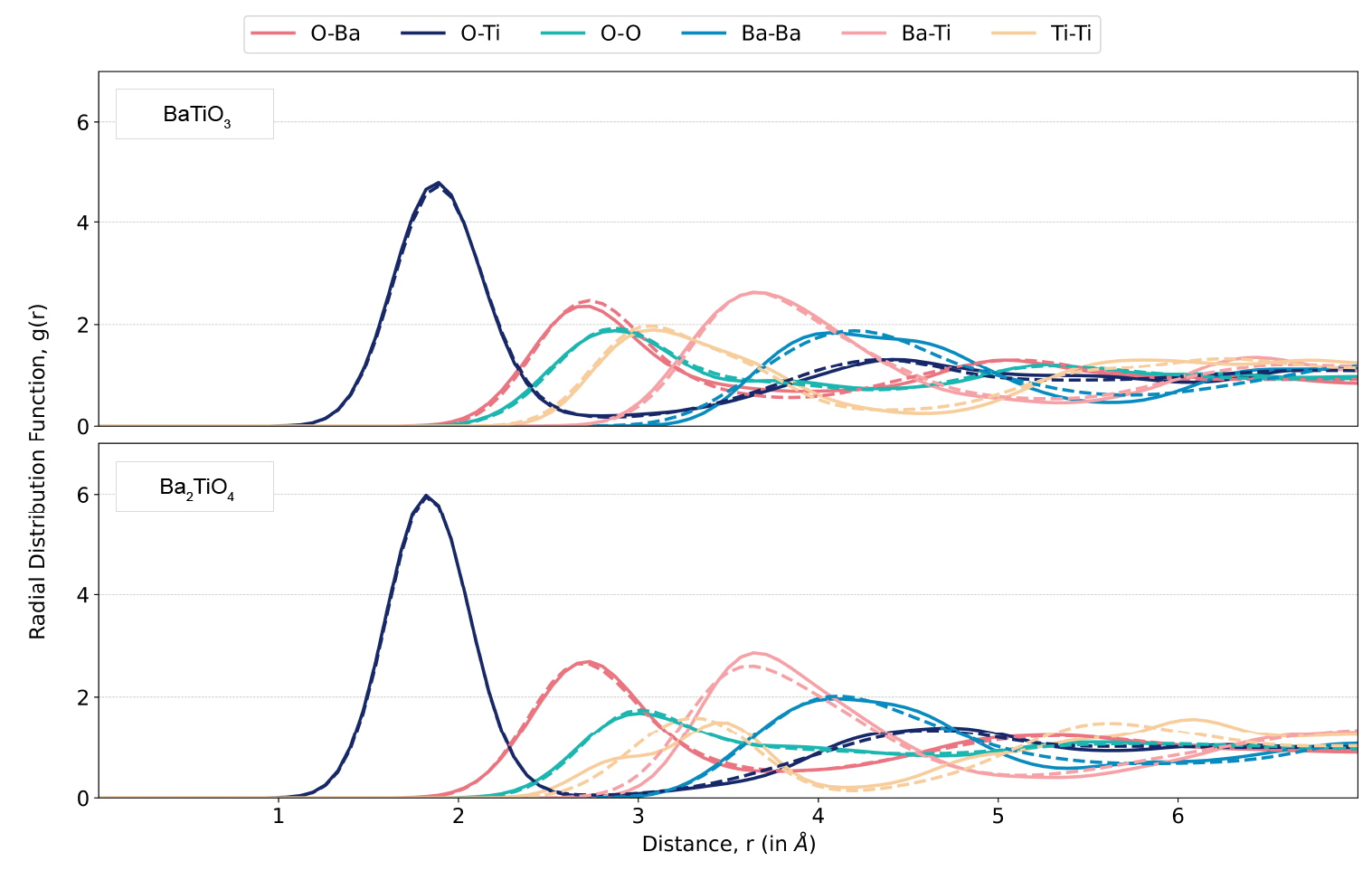}
    \caption{Comparison of RDFs obtained from ACE-MD for BaTiO3 (top) and Ba2TiO4 (bottom), shown with the solid line, with corresponding RDFs obtained from AIMD, shown with the dashed line.}
    \label{fig:rdf_benchmark}
\end{figure}

To ensure spatial correlations are properly captured, we run ACE-MD on MPMorph equilibrated structures for all compositions that were used for training the potential, and the radial distribution functions (RDFs) are computed for all pairs of atomic specie and compared to those obtained from AIMD. The RDFs for BaTiO3 and Ba2TiO4 at 1000K are shown in Figure \ref{fig:rdf_benchmark}. As can be seen, the model recovers the correct short range configuration, with the position and magnitude of the first peak (i.e., the first shell) of the ACE-MD trajectory lining up almost exactly with the AIMD trajectory. There is however, a deviation after the first shell, notably for radial distances > 5$\mathring{A}$. This deviation can be explained by the neighborhood lists used during ACE-MD only considering neighbors within 5$\mathring{A}$, as well as the general lack of medium and long range order in these systems. 

\newpage
\section{Adapting Onsager transport framework to diffusive transport in amorphous solids}

The Onsager transport coefficients are typically defined with respect to a reference specie in the system \cite{fong2020onsager}. For electrolytes, this is often the solvent being used. Due to the lack of such a reference specie, in our case we use the center of mass of the system as the reference, defining all displacements and velocities in the frame of the center of mass when computing correlation functions. This leads to an over-determined system: not all coefficients in the transport matrix are independent. By the Onsager reciprocity theorem and the second law of thermodynamics, the matrix is symmetric and positive semi-definite respectively. Further, since the reference itself depends on all other species in the system, every row and column in the transport matrix has one coefficient which is the linear combination of all other transport coefficients of that row or column:
\begin{equation}
\label{eqn:Lij_dependence}
    \Sigma_i L_{ij} = 0
\end{equation}
This equation holds due to $L_{ij}$ being computed in the center-of-mass frame of reference, which leads to zero net mass flux in the system. For brevity, we left out two crucial constraints to the flux formulation in Equation 1 of the manuscript: linear irreversible thermodynamics assumes the presence of local thermodynamic equilibrium, and the condition of electro-neutrality. These can be stated as follows:
\begin{equation}
\label{eqn:local_equil}
    \Sigma_i c_i \nabla {\mu}_i = 0
\end{equation}

\begin{equation}
\label{eqn:charge_balance}
    \Sigma_i q_i c_i = 0
\end{equation}
Here, $q_i$ and $c_i$ the charge and concentration of specie $i$, and $\nabla {\mu}_i$ is the chemical potential gradient acting on specie $i$. It can be easily shown that Equations \ref{eqn:Lij_dependence}, \ref{eqn:charge_balance} and \ref{eqn:local_equil} constrain one variable of the system, i.e., one mass flux. These constraints allow us to exclude the O-ion flux from the definition of $K_D$, since this term is fixed by the definition of the Ba and Ti fluxes. We essentially consider a linear combination of the fluxes when defining the effective rate constant ($K_D$), thereby allowing us to capture the effect of changing driving forces and transport properties on all ions across the interface.
\begin{figure}
    \centering
    \includegraphics[width=\linewidth]{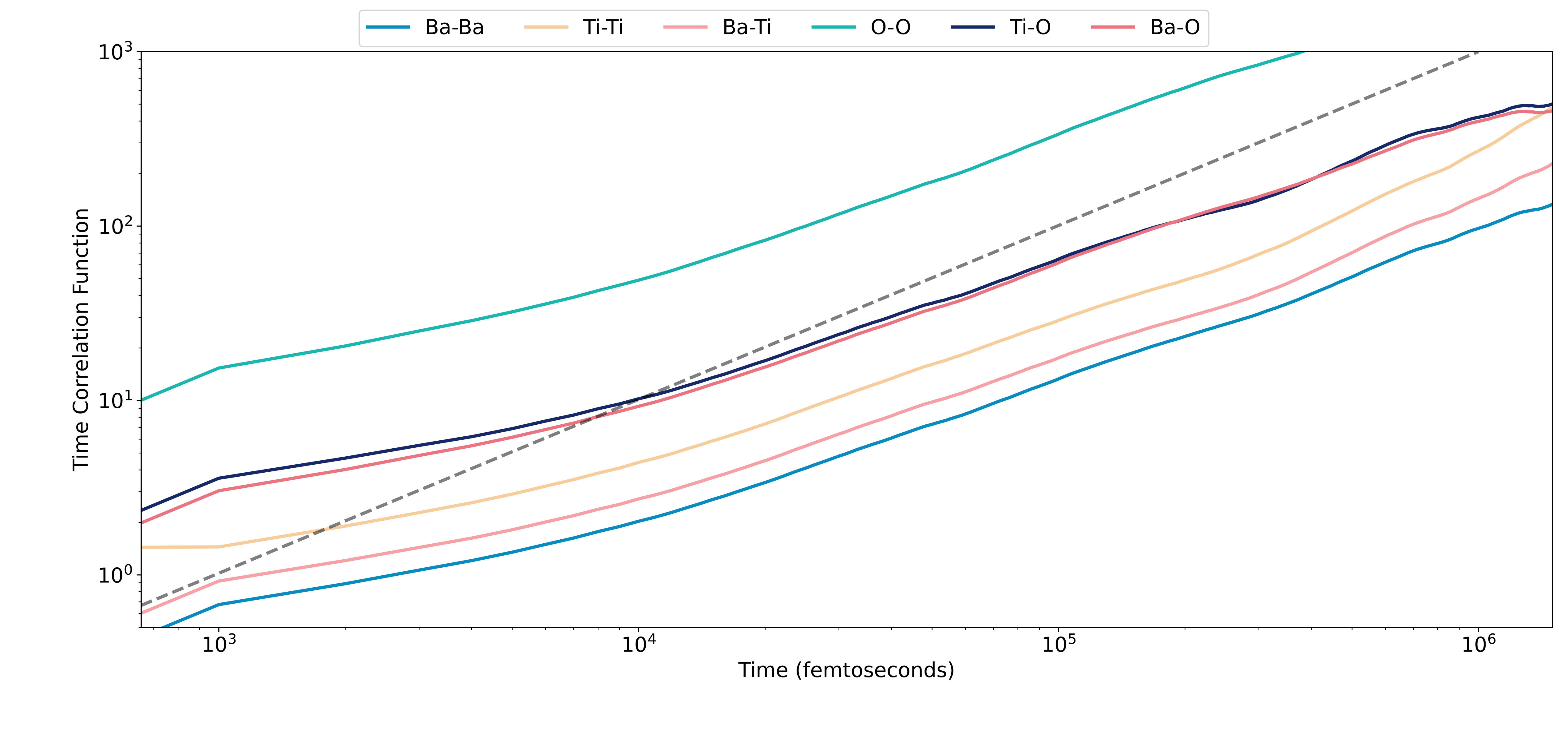}
    \caption{Displacement time-correlation function vs time plot for an ACE-MD trajectory of \ce{BaTiO3} at 750K in the log-log scale. A dashed line showing a correlation function with perfectly linear slope has been provided as a guide to the eye.}
    \label{fig:fitting}
\end{figure}
In this work, we have three species with a first coordination shell within 3.5 $A^o$ of every atom, allowing us to calculate the ionic transport using  `smaller' systems of approximately 100 atoms. Compared to traditional self-diffusion coefficients, the Onsager transport matrix requires more statistics to fit robustly. This is primarily due to the lack of sufficient cross-ion correlation statistics (i.e., correlated movement between ions) for short timescales. Figure \ref{fig:fitting} shows the displacement time correlation function as a function of time in an ACE-MD trajectory of \ce{BaTiO3} at 750K during the first 2 nanoseconds in log scale.  The time scale typically achievable with AIMD would be between 10-50 picoseconds ($1\times10^4-5\times10^4$), where the system has not yet reached the diffusive regime (notice the difference in slopes between the computed time-correlation functions and the linear curve, shown with a dashed line). After 1.8 nanoseconds, a linear curve fit becomes feasible for all correlation functions, as their slopes align closely with the linear trend. Additional windows of linearity emerge when the simulation is extended, particularly for MD runs conducted at higher temperatures. Overall, using ACE-MD (or any MLIP-based MD) is essential for accurately fitting the full Onsager transport matrix in such ionic systems. It is worth noting that the quality of the fits for cross-coefficients is generally lower than for the self-terms. This issue could be mitigated by running ACE-MD simulations for longer durations or on larger systems. For this study, we ensured that the uncertainty in the fit was no greater than 40\% for the cross-transport terms and no greater than 20\% for the net and self-transport terms. Trajectories failing to meet this criterion were excluded. Reliable fits for cross-ion coefficients were not achievable for certain phases at 750 K. Consequently, we opted to exclude data from 750 K in our analysis. The code-base for the transport coefficients can be found at \texttt{https://github.com/vir-k01/py-OATS}, and will be made public on publication of the manuscript.

\section{Chemical Potential Diagram for Ba-Ti-O system}

\begin{figure}[h!]
    \centering
    \includegraphics[width=0.8\linewidth]{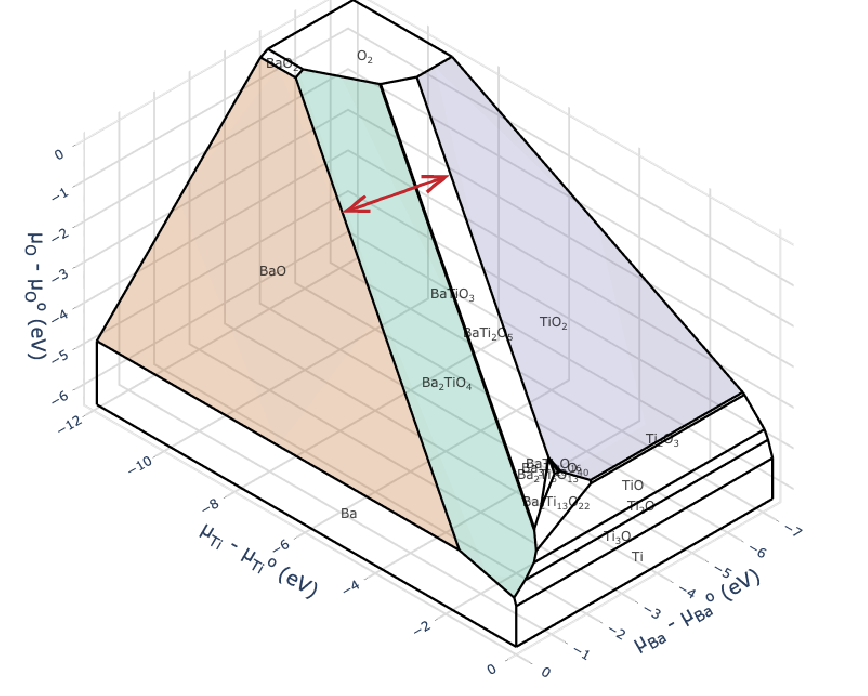}
    \caption{Chemical potential diagram for the Ba-Ti-O chemical system at 0K.}
    \label{fig:chempot}
\end{figure}

The chemical potential diagram at 0K as computed using \texttt{pymatgen} with all entries from the Materials Project in the Ba-Ti-O chemical space, is shown in Figure \ref{fig:chempot}. In a reaction involving \ce{BaO} and \ce{TiO2}, the corresponding planes are shaded red and purple, respectively. The driving force for ion movement across the reaction interface is represented by the shortest distance between these two planes, approximately indicated by the red arrow. Conversely, for a reaction between \ce{BaO} and \ce{Ba2TiO4} (shaded green), the two planes share an edge, resulting in a shortest distance of zero. This indicates no net flux across the interface, signifying thermodynamic equilibrium. For details on how these diagrams are computed, refer to Ref \cite{yokokawa1999generalized, neilson_modernist_2023}. 

\section{Derivation of rate across a spherical powder interface}
\label{sec:rate_derivation}

\begin{figure}[h!]
    \centering
    \includegraphics[width=0.5\linewidth]{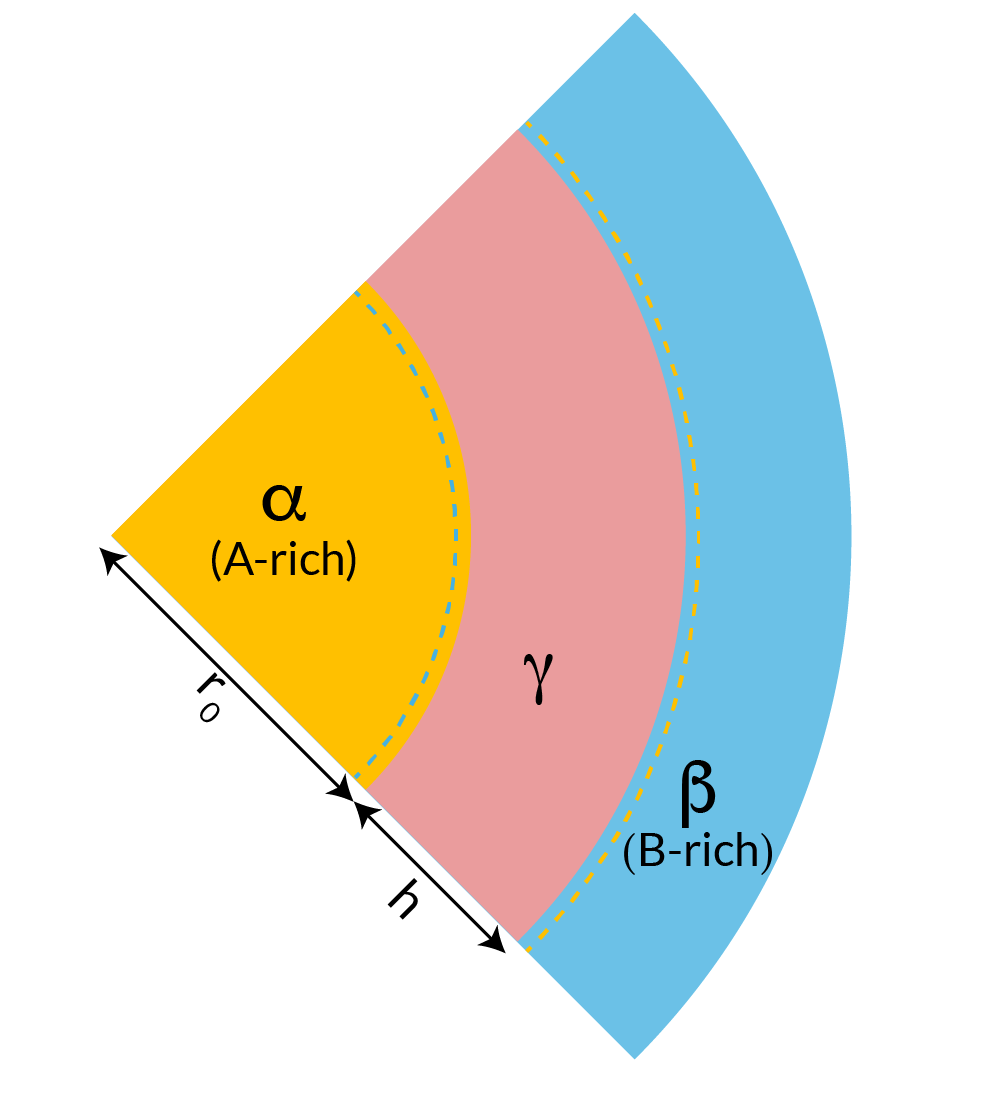}
    \caption{A schematic for the core-shell model for diffusion-controlled growth, wherein diffusion of A from $\alpha$ to $\beta$ or B from $\beta$ to $\alpha$ controls the rate of growth of $\gamma$. $r_0$ is the radius of the (spherical) particle of reactant $\alpha$ and $h$ is the instantaneous thickness of the $\gamma$ product.}
    \label{fig:core-shell}
\end{figure}

Following \cite{dheurle_theoretical_1995}, we start with the case for diffusion-limited growth in the geometry shown in Figure \ref{fig:core-shell}, wherein we growth of $\gamma$ happens onto a reactant particle of $\alpha$. The growth rate of the product layer $\gamma$ (=$\frac{dh}{dt}$) can be expressed as:
\begin{equation*}
    \frac{dh}{dt} = V^{\gamma}\Sigma_i \frac{|J_i^{\gamma}|}{n_i^\gamma}
\end{equation*}

Here, $J_i^\gamma$ is the mass flux of the $i^{th}$ ion across $\gamma$, whose transport from one reactant to the other is needed for continual growth of $\gamma$. We use the fact that transport of one mole of $i$ leads to the formation of $\frac{V_i^\gamma}{n_i^\gamma}$ moles of the $\gamma$ phase, where $V_i^\gamma$ and $n_i^\gamma$ are the molar volumes and molar fractions of specie $i$ in $\gamma$. Using our definition of fluxes and the effective diffusion rate constant (Equations 1, 7 of the manuscript), we can rewrite this expression as:
\begin{equation*}
    \frac{dh}{dt} = \frac{K_D}{h}
\end{equation*}

Assuming $K_D$ is independent of $h$, we can solve this equation using the initial condition of $h (t=0) =0$:
\begin{equation*}
    h(t) = \sqrt{2K_Dt}
\end{equation*}

If we now define the ``fraction of reactants consumed''($y(t)$) as the change in volume of the reactant $\alpha$ (from Figure \ref{fig:core-shell}):
\begin{equation*}
    y = \frac{V_\gamma}{V_\alpha} = 1 - [\frac{(r_0 - h(t))}{r_0}]^3
\end{equation*}

Rearranging this expression using the solution for $h(t)$ gives:
\begin{equation*}
    [1 - (1-y(t))^{1/3}]^2 = \frac{2K_Dt}{r_0}
\end{equation*}

Obviously, on repeating this exercise for the case where $\gamma$ grows onto a particle of $\beta$ instead, the only difference is $r_0$ becomes the size of the reactant $\beta$. This equivalence is because transport of all species (in this 3-component case) are inter-dependent through the Onsager definition of the flux, and we assume $K_D$ does not depend on $h$ or time. 

\newpage
\section{Computed Onsager transport coefficients}

\begin{figure}[h!]
    \centering
    \includegraphics[width=\linewidth]{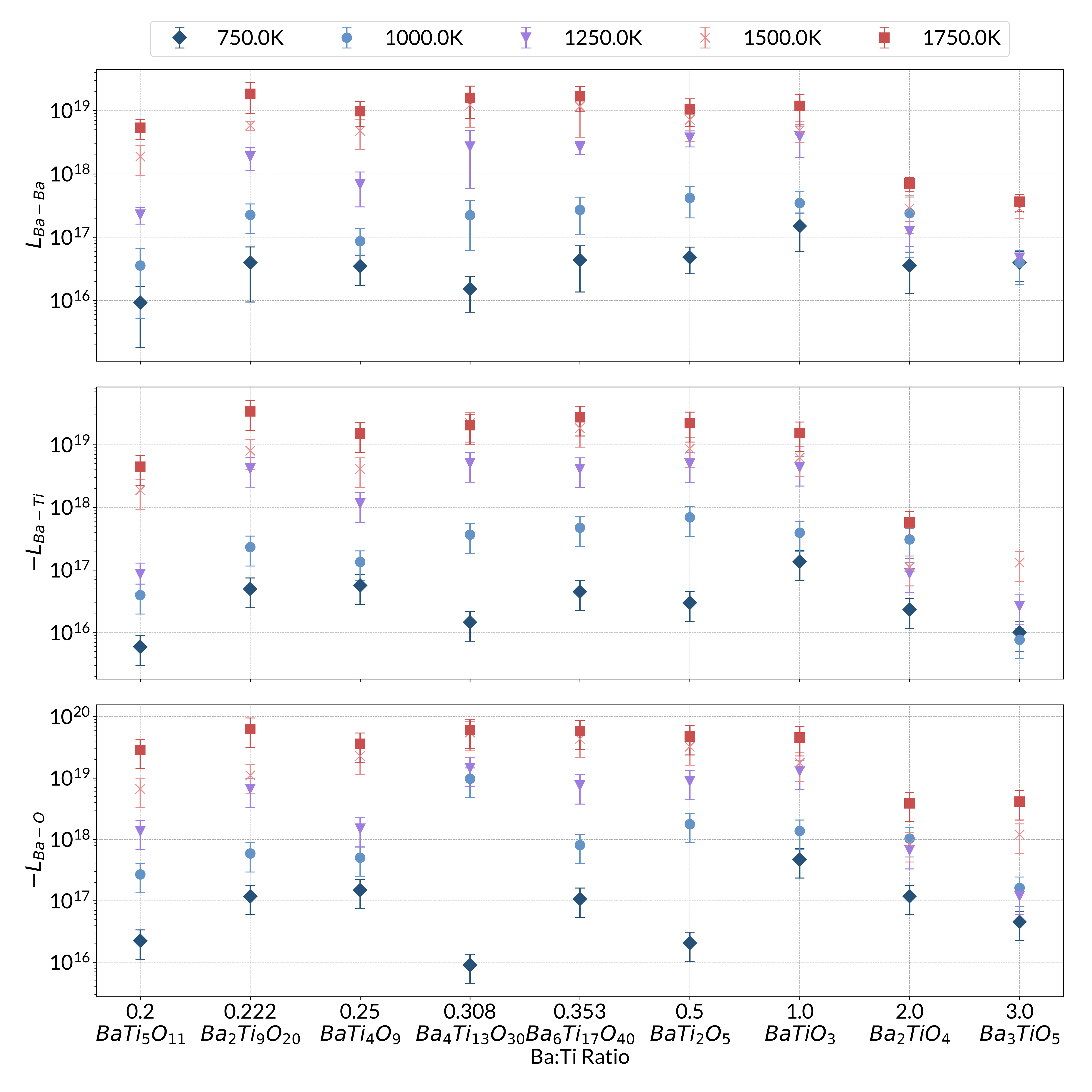}
    \caption{The computed Onsager transport coefficients (in units of $1/(eV-cm^3-sec)$) for different Ba-Ti-O phases, ordered by increasing Ba:Ti ratio, corresponding to the Ba-Ba correlation (top), Ba-Ti correlation (middle) and Ba-O correlation (bottom).}
    \label{fig:Lij_BTO_1}
\end{figure}

\begin{figure}[h!]
    \centering
    \includegraphics[width=\linewidth]{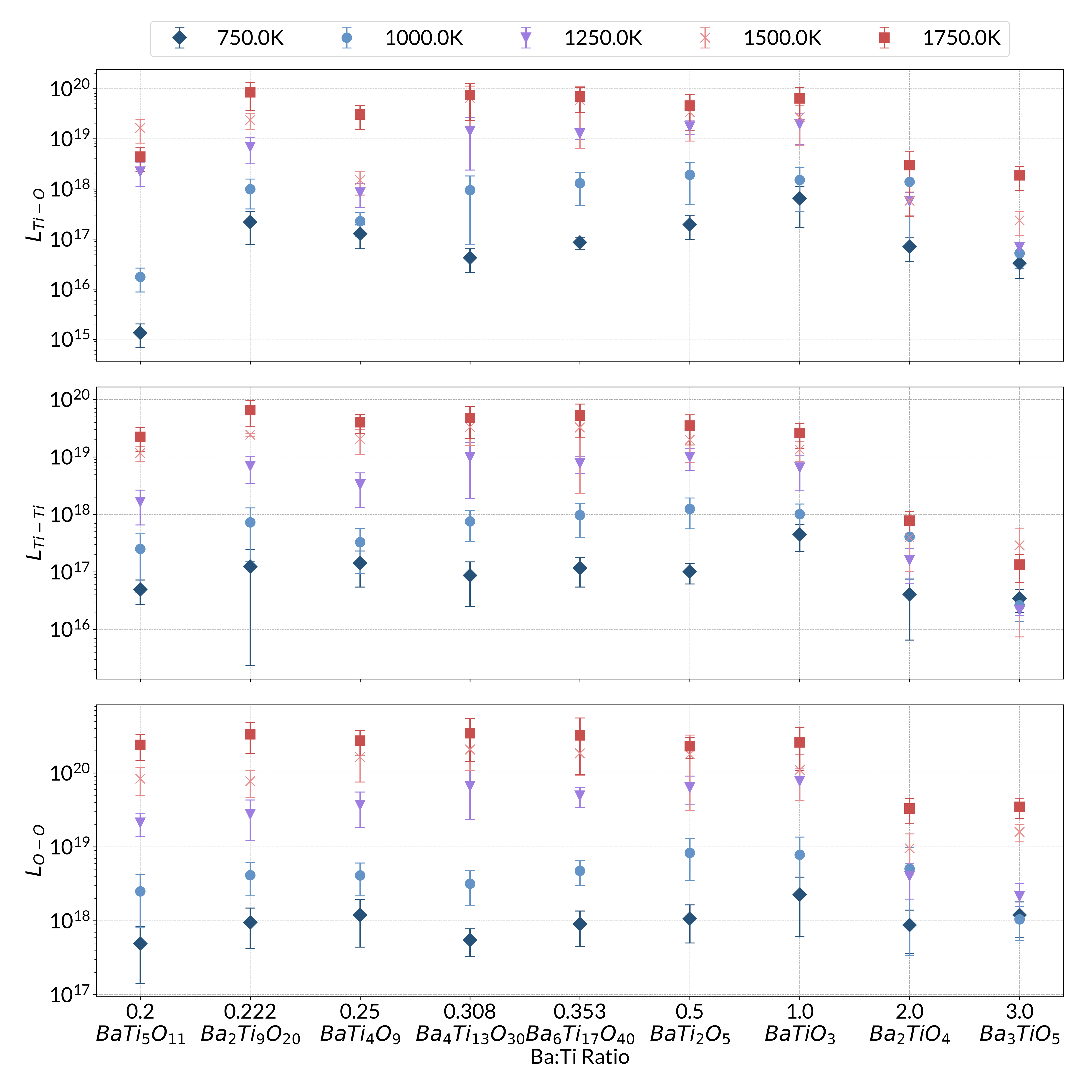}
    \caption{The computed Onsager transport coefficients (in units of $1/(eV-cm^3-sec)$) for different Ba-Ti-O phases, ordered by increasing Ba:Ti ratio, corresponding to the Ti-O correlation (top), Ti-Ti correlation (middle) and O-O correlation (bottom).}
    \label{fig:Lij_BTO_2}
\end{figure}

\newpage

\section{ReactCA simulations without cross-ion transport effects}
\begin{figure}[h!]
    \centering
    \includegraphics[width=\linewidth]{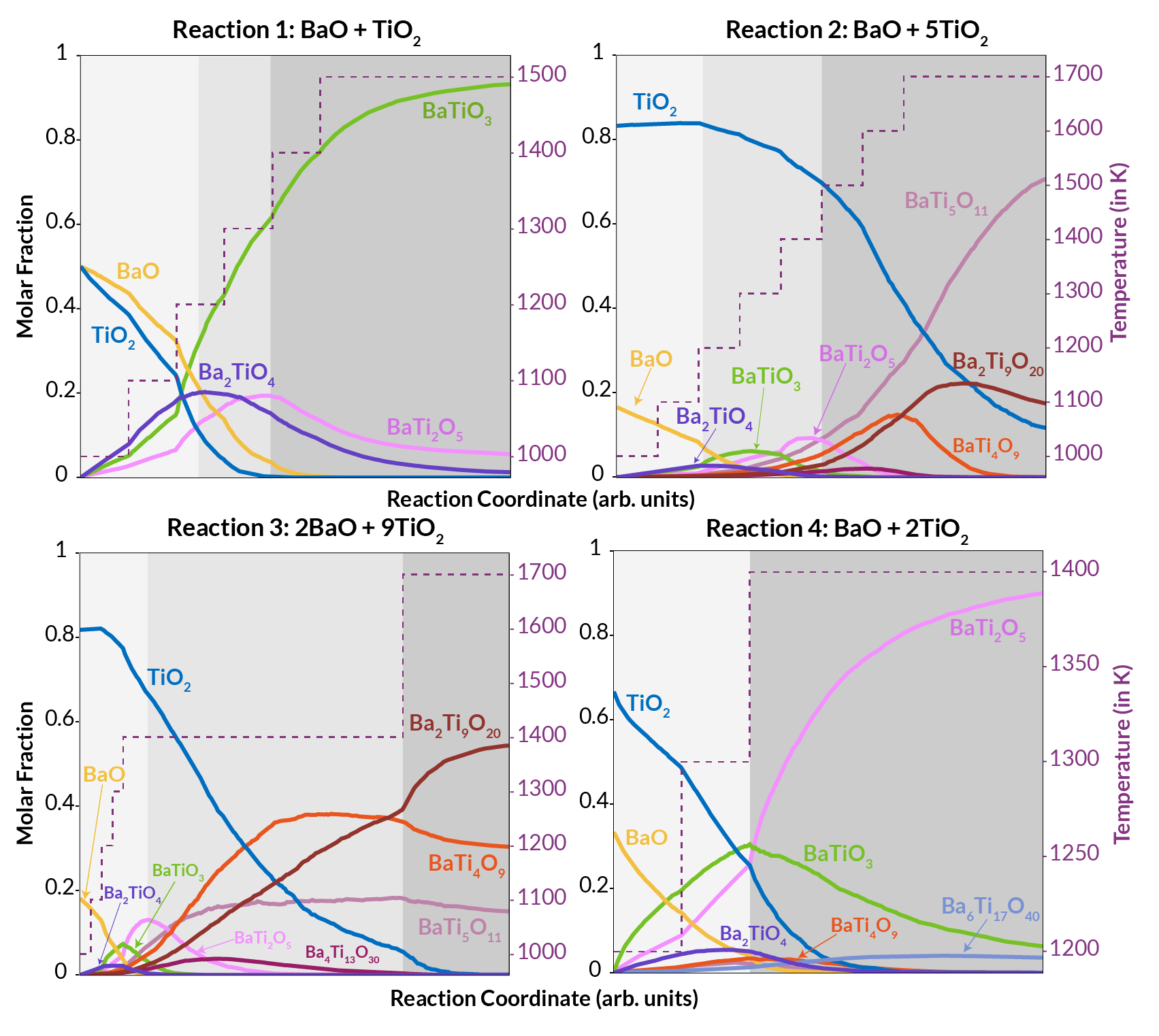}
    \caption{ReactCA simulation incorporating only self-diffusion based transport fluxes into the scoring function.}
    \label{fig:bto_with_self}
\end{figure}

Reactions 1-4, were simulated using only the ``self'' transport fluxes in the scoring function. The scoring function for this case is given as:
\begin{equation}
\label{eqn:score_self}
\begin{split}
    & S = \sigma_1(\frac{K_{D}}{r_0^2s} * \frac{\Delta G^*}{k_BT}) * \sigma_2 (\frac{T}{T_{m, reactant}}) \\
    & K_D = \Sigma_j \frac{|L^{self, \gamma}_{jj} V^{\gamma}\cdot \text{min}(\mu_j^\alpha - \mu_j^\beta)|}{n^{\gamma}_j} \\
\end{split}
\end{equation}
Here, $L_{jj}^{\text{self}}$ can be estimated from the self-diffusion coefficient of specie $j$ ($D_{j}$) and the concentration of $j$ ($c_{j}$) in the product layer at temperature T as: 
\begin{equation}
    L_{jj}^{\text{self}} = \frac{D_{j}c_{j}}{k_BT}
\end{equation}
Figure \ref{fig:bto_with_self} shows the results of the simulations. The final product distributions are qualitatively very similar to the simulations from the manuscript which utilized the full transport matrix with cross-effects in the scoring function.

\section{Nucleation barrier estimation using PIRO}
In order to consider nucleation kinetics into our score, we also experimented with the approach of using lattice matching/structural similarity between precursors and solid-state products as a proxy for the degree of heterogeneous nucleation as described in the PIRO approach. Table \ref{tab:reaction-g-j} shows the estimated nucleation barrier ($\Delta G^*$) and rate of nucleation ($J_{\text{nuc}} \equiv \exp{\frac{-\Delta G^*}{k_BT}}$) for the formation of the products considered in this study from three precursor combinations: \ce{BaO} + \ce{TiO2}, \ce{BaO2} + \ce{TiO2}, \ce{BaCO3} + \ce{TiO2}.

\begin{table}[h!]
\centering
\begin{tabular}{|l|l|l|}
\hline
\textbf{Reaction} & \textbf{$\Delta G*$} & \textbf{$J_{\text{nuc}}$} \\ \hline
\ce{BaO + TiO2 -> BaTiO3} & 0.0435 & \textbf{0.713977} \\ 
\ce{BaO + TiO2 -> Ba2TiO4} & 1.0888 & \textbf{0.0002} \\ 
\ce{BaO + TiO2 -> Ba6Ti17O40} & 6.2668 & 0.0 \\ 
\ce{BaO + TiO2 -> Ba4Ti13O30} & 6.7435 & 0.0 \\ 
\ce{BaO + TiO2 -> Ba2Ti9O20} & 8.7711 & 0.0 \\ 
\ce{BaO + TiO2 -> BaTi2O5} & 0.3663 & \textbf{0.0588} \\ 
\ce{BaO + TiO2 -> Ba3TiO5} & 1.8074 & 0.0 \\ 
\ce{BaO + TiO2 -> BaTi4O9} & 1.5753 & 0.0 \\ 
\ce{BaO + TiO2 -> BaTi5O11} & 5.8769 & 0.0 \\ 
\ce{BaO + TiO2 -> BaTi6O13} & 2.3497 & 0.0 \\ 
\ce{BaO2 + TiO2 -> BaTiO3 + O2} & 0.5409 & 0.0152 \\ 
\ce{BaO2 + TiO2 -> Ba2TiO4 + O2} & 99.9718 & 0.0 \\ 
\ce{BaO2 + TiO2 -> Ba6Ti17O40 + O2} & 39.3189 & 0.0 \\ 
\ce{BaO2 + TiO2 -> Ba4Ti13O30 + O2} & 34.8828 & 0.0 \\ 
\ce{BaO2 + TiO2 -> Ba2Ti9O20 + O2} & 48.8114 & 0.0 \\ 
\ce{BaO2 + TiO2 -> BaTi2O5 + O2} & 2.1875 & 0.0 \\ 
\ce{BaO2 + TiO2 -> Ba3TiO5 + O2} & 26.5411 & 0.0 \\ 
\ce{BaO2 + TiO2 -> BaTi4O9 + O2} & 9.6281 & 0.0 \\ 
\ce{BaO2 + TiO2 -> BaTi5O11 + O2} & 33.4494 & 0.0 \\ 
\ce{BaO2 + TiO2 -> BaTi6O13 + O2} & 16.2007 & 0.0 \\ 
\ce{BaCO3 + TiO2 -> BaTiO3 + CO2} & 0.1441 & 0.328 \\ 
\ce{BaCO3 + TiO2 -> Ba2TiO4 + CO2} & 1.7433 & 0.0 \\ 
\ce{BaCO3 + TiO2 -> Ba6Ti17O40 + CO2} & 28.4083 & 0.0 \\ 
\ce{BaCO3 + TiO2 -> Ba4Ti13O30 + CO2} & 44.2047 & 0.0 \\ 
\ce{BaCO3 + TiO2 -> Ba2Ti9O20 + CO2} & 39.2851 & 0.0 \\ 
\ce{BaCO3 + TiO2 -> BaTi2O5 + CO2} & 2.3927 & 0.0 \\ 
\ce{BaCO3 + TiO2 -> Ba3TiO5 + CO2} & 0.8953 & 0.001 \\ 
\ce{BaCO3 + TiO2 -> BaTi4O9 + CO2} & 6.1428 & 0.0 \\ 
\ce{BaCO3 + TiO2 -> BaTi5O11 + CO2} & 23.7557 & 0.0 \\ 
\ce{BaCO3 + TiO2 -> BaTi6O13 + CO2} & 6.1576 & 0.0 \\ \hline
\end{tabular}
\caption{$\Delta G^*$ values, and $J_{\text{nuc}}$ computed at 1500K using the PIRO approach.}
\label{tab:reaction-g-j}
\end{table}

\pagebreak

\end{document}